\documentclass[12pt]{iopart}

\usepackage{iopams}

\usepackage{graphicx}   
\usepackage{dcolumn}    
\usepackage{bm}         
\usepackage{multirow}   

\eqnobysec  

\begin{document}

\topical[Slip effects in polymer thin films]{Slip effects in
polymer thin films}

\author{O B\"aumchen and K Jacobs}

\address{Experimental Physics, Saarland University, Campus, D-66123 Saarbr\"ucken, Germany}
\ead{k.jacobs@physik.uni-saarland.de}

\begin{abstract}
Probing the fluid dynamics of thin films is an excellent tool to
study the solid/liquid boundary condition. There is no need for
external stimulation or pumping of the liquid due to the fact that
the dewetting process, an internal mechanism, acts as a driving
force for liquid flow. Viscous dissipation within the liquid and
slippage balance interfacial forces. Thereby, friction at the
solid/liquid interface plays a key role towards the flow dynamics of
the liquid. Probing the temporal and spatial evolution of growing
holes or retracting straight fronts gives, in combination with
theoretical models, information of the liquid flow field and
especially the boundary condition at the interface. We review the
basic models and experimental results obtained during the last years
with exclusive regard to polymers as ideal model liquids for fluid
flow. Moreover, concepts that aim on explaining slippage on the
molecular scale are summarized and discussed.
\end{abstract}



\submitto{\JPCM}


\maketitle

\section{Introduction}
\label{Intro} Understanding liquid flow in confined geometries
plays a huge role in the field of micro- and nanofluidics
\cite{Squ05}. Nowadays, microfluidic or so-called lab-on-chip
devices are utilized in a wide range of applications. Pure
chemical reactions as well as biological analysis performed on
such a microfluidic chip allow a high performance while solely
small amounts of chemicals are needed. Thereby, analogies to
electronic large-scale integrated circuits are evident. Thorsen
\etal fabricated a microfluidic chip with a high density of
micromechanical valves and hundreds of individually addressable
chambers \cite{Qua02}. Recent developments tend to avoid huge
external features such as pumps to control the flow by designing
analogues to capacitors, resistors or diodes that are capable to
control currents in electronic circuits \cite{Les09}.

By reducing the spacial dimensions of liquid volume in confined
geometries, slippage can have a huge impact on flow dynamics.
Especially the problem of driving small amounts of liquid volume
through narrow channels has drawn the attention of many researchers
on slip effects at the solid/liquid interface. The aim is to reduce
the pressure that is needed to induce and to maintain the flow.
Hence, the liquid throughput is increased and, what is important in
case of polydisperse liquids or mixtures, a low dispersity according
to lower velocity gradients perpendicular to the flow direction is
generated. Confined geometries are realized in various types of
experiments: The physics and chemistry of the imbibition of liquids
by porous media is of fundamental interest and enormous
technological relevance \cite{Bea72}. Channel-like three-dimensional
structures can be used to artificially model the situation of fluids
in confinement. Moreover, the small gap between a colloidal probe
and a surface filled with a liquid (e.g realized in surface force
apparatus or colloidal probe atomic force microscopy experiments) is
a common tool to study liquid flow properties. Besides the
aforementioned experimental systems, knowledge in preparation of
thin polymer films has been extensively gained due to its enormous
relevance in coating and semiconductor processing technology. Such a
homogeneous nanometric polymer film supported by a very smooth
substrate, as for example a piece of a silicon (Si) wafer, exhibits
two relevant interfaces, the liquid/air and the substrate/liquid
interface. Si wafers are often used according to their very low
roughness and controllable oxide layer thickness. Yet, also highly
viscous and elastically deformable supports, such as e.g.
polydimethylsiloxane (PDMS) layers, are versatile substrates.

\begin{figure}[t]
\begin{center}
\includegraphics[width=0.6\textwidth]{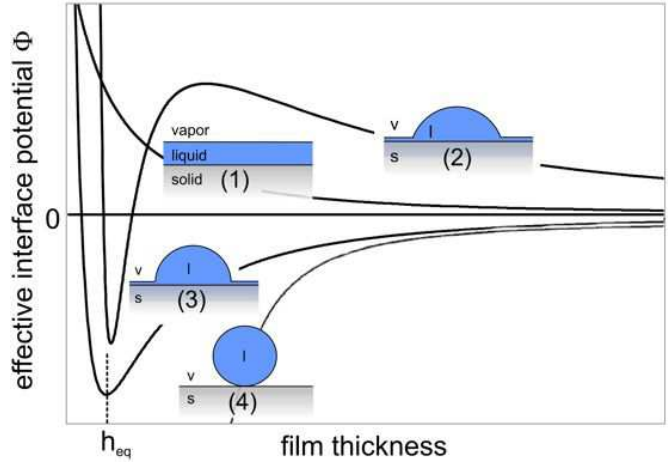}
\end{center}
\caption{Different shapes of the effective interface potential
$\phi(h)$ associated with different wetting conditions. Curve (1)
characterizes a \textit{stable} liquid film. Curve (2) represents a
\textit{metastable}, curve (3) and (4) an \textit{unstable}
situation.} \label{graphphiskizze}
\end{figure}

The stability of a thin film is governed by the effective interface
potential $\phi$ as function of film thickness $h$. In case of
dielectric systems, $\phi(h)$ is composed of an attractive
van-der-Waals part and a repulsive part
\cite{Vri66,Her98,See012,Jac08}. For the description of
van-der-Waals forces of a composite substrate, the layer thicknesses
and their respective polarisation properties have to be taken into
account \cite{See013}. Thereby, three major situations of a thin
liquid film have to be distinguished: \textit{stable},
\textit{unstable} and \textit{metastable} films. As illustrated in
Fig. \ref{graphphiskizze} by the typical curves of $\phi(h)$, a
\textit{stable} liquid film is obtained if the effective interface
potential is positive and monotonically decaying (cf. curve (1) in
Fig. \ref{graphphiskizze}). The equilibrium film thickness $h_{eq}$
is infinite and the liquid perfectly wets the substrate. In case of
a global minimum of $\phi(h)$, curves (2) and (3) in Fig.
\ref{graphphiskizze}, the system can minimize its energy and a
finite value for $h_{eq}$ results. The \textit{metastable} situation
is furthermore characterized by a potential barrier that the system
has to overcome to reduce its potential energy (cf. curve (2) in
Fig. \ref{graphphiskizze}). Curve (3) and (4) characterize
\textit{unstable} conditions, since every slight fluctuation in film
height will drive the system towards the global minimum. The
wettability of the substrate by the liquid is correlated to the
depth of the minimum at $\phi(h_{eq})$. The deeper the global
minimum of $\phi$, the larger is the equilibrium contact angle of
the liquid on the surface. A nearly 180$\,^{\circ}$ situation is
depicted in curve (4) of Fig. \ref{graphphiskizze}.

\begin{figure}[t]
\begin{center}
\includegraphics[width=\textwidth]{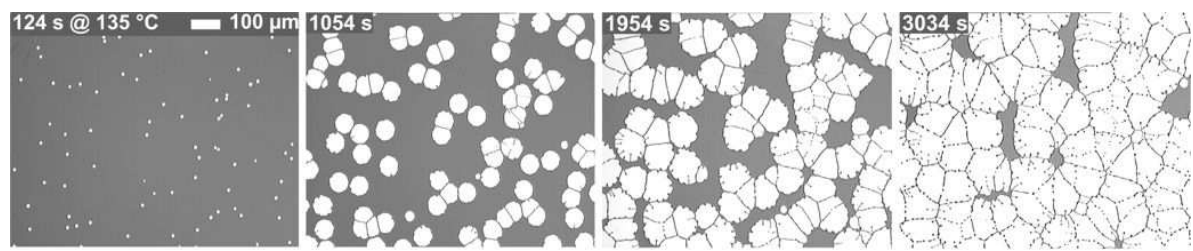}
\end{center}
\caption{Dewetting of a 80\,nm PS(65\,kg/mol) film at
$T=135\,^{\circ}$C from a hydrophobized Si substrate captured by
optical microscopy (adapted from \cite{Jac00}).}
\label{graphentnetzung}
\end{figure}

For a 100\,nm polystyrene (PS) film on a hydrophobized Si wafer with
a native oxide layer, dewetting starts after heating the sample
above the glass transition temperature of the polymer. Holes
nucleate according to thermal activation or nucleation spots (dust
particles, inhomogeneities of the substrate or of the polymer film)
and grow with time, cf. Fig. \ref{graphentnetzung}. The subsequent
stages of dewetting are characterized by the formation of liquid
ridges by coalescence of growing holes and traveling fronts. Very
thin films in the range of several nanometers may become unstable
and can dewet according to thermally induced capillary modes, that
are amplified by forces contributing to the effective interface
potential. This phenomenon is characterized by the occurrence of a
preferred wavelength and is called \textit{spinodal dewetting}. To
study thin film flow with regard to the influence of slippage,
especially nucleated holes enable an easy experimental access for
temporal and spatial observation.

While fronts retract from the substrate and holes grow, a liquid rim
is formed at the three-phase contact line due to conservation of
liquid volume. A common phenomenon is the formation of liquid bulges
and so-called ''fingers'' due to the fact that the liquid rim
becomes unstable, similar to the instability of a cylindrical liquid
jet that beads up into droplets (e.g. in case of a water tap). This
so-called \textit{Rayleigh-Plateau instability} is based on the fact
that certain modes of fluctuations become amplified and surface
corrugations of a characteristic wavelength become visible. If two
holes coalesce, a common ridge builds up that in the end decays into
single droplets due to the same mechanism. The final stage is given
by an equilibrium configuration of liquid droplets arranged on the
substrate exhibiting a static contact angle. Actually, the final
state would be one single droplet, since the Laplace pressure in
droplets of different size is varying. Yet then, a substantial
material transport must take place, either over the gas phase or via
an equilibrium film between the droplets. This phenomenon is also
called \textit{Ostwald ripening}. In polymeric liquids, these
transport pathways are usually extremely slow so that a network-like
pattern of liquid droplets is already termed ''final stage''.
Besides the dynamics and stability of thin liquid films driven by
intermolecular forces, a recent article by Craster and Matar reviews
further aspects such as e.g thermally or surfactant driven flows
\cite{Cra09}.

In the first part of this topical review, basic concepts from
hydrodynamic theories in the bulk situation to corresponding models
with regard to confinement are introduced. Polymers are regarded as
ideal model liquids due to their low vapor pressure, the available
chemical pureness and furthermore the fact that their viscosity can
be controlled in a very reliable manner. By setting the viscosity
via temperature, the experimental conditions can be tuned so that
dewetting dynamics can be easily captured. Mass conservation can
safely be assumed, which consequently simplifies the theoretical
description. The dewetting dynamics governed by the driving forces
and the mechanisms of energy dissipation will be discussed also with
regard to the shape of liquid ridges. The second part summarizes
experimental studies concerning the dynamics of two different
dewetting geometries: the straight front geometry and the growth of
holes. Especially the influence of parameters such as dewetting
temperature, viscosity and molecular weight of the polymer will be
discussed in detail. Moreover, we focus on various scenarios at the
solid/liquid interface on the molecular level. Simulations such as
for example molecular-dynamic (MD) studies help to obtain more
information and are supportive to gain insight into the molecular
mechanism of slippage.

\section{Basic theoretical concepts}

In this section we aim to describe the main concepts of fluid
dynamics especially in a confined geometry. Depending on the type of
liquid, viscous or even viscoelastic effects have to be considered
and deviations from Newtonian behavior might become non-negligible.
Since we concentrate in this article on polymer melts, viscosity and
viscoelasticity can be varied by chain length and branching of the
polymer. An important aspect of a moving liquid is the velocity
profile.

\subsection{Polymer properties}
\label{properties} For a comprehensive understanding of slippage,
some important polymer properties like glass transition temperature,
viscosity and viscoelasticity have to be taken into account.
Especially in geometries like in a liquid film, confinement effects
are a concern. A detailed description can be found in textbooks
\cite{Rub03,Jon99}.

\subsubsection{Polymer physics}
\label{polymers}

Polymers are synthesized by polymerization of monomers of molar mass
$M_{mono}$. To characterize the polydispersity of polymer chains in
a solution or in a melt, the polydispersity index $M_w/M_n$ is
calculated as the ratio of mean values given by the weight $M_w$ and
number averaged $M_n$ molecular weight. Polymers are able to change
their conformations. The radius of gyration characterizes the
spatial dimension of the polymer and is given as the mean square
displacement between monomers and the polymer's center of mass. In
an isotropic configuration, the shape of the polymer chain can be
approximated as a spherical entity. Polystyrene, abbreviated PS,
which is commonly used as a melt in dewetting experiments, is a
linear homopolymer with $M_{mono}=104$\,g/mol. Besides other
properties concerning the micro-structure of polymer chains such as
the tacticity and the architecture (linear, branched, ring-shaped),
physical properties are of special interest. If a melt is heated
above its glass transition temperature $T_g$, a phase transition
from the glassy phase to the liquid phase occurs and the polymer
becomes liquid. Randomly structured macromolecules such as atactic
polymers avoid the formation of semi-crystalline domains below $T_g$
and exhibit a pure amorphous phase. The glass transition of a bulk
polymer or of a polymeric thin film can be observed e.g. via probing
its linear expansion coefficient. Although, the glass transition is
based on a kinematic effect and does not occur due to a
rearrangement process of polymer chains. Therefore, the glass
transition usually takes place on a specific temperature range and
not at a exactly allocable temperature. According to the increased
mobility of shorter chains, their glass transition temperature is
decreased significantly. For PS with a sufficiently large
chain-length, $T_g$=100\,$^\circ$C.

The viscosity $\eta$ of a polymer melt measures the inner friction
of polymer chains and governs the time scale of flow processes. Due
to the fact that the mobility of chains increases while the
temperature increases, the viscosity decreases. Internal stresses
relax and dynamical processes proceed faster. The most common
description of the functional dependency of the viscosity $\eta$ of
a polymer on temperature $T$ was developed by Williams, Landel and
Ferry:

\begin{equation}
\label{WLF} \eta=\eta_g\exp{\frac{B(T_g-T)}{f_g(T-T_\infty)}}.
\end{equation}

\noindent In this so-called WLF-equation, $\eta_g$ denotes the
viscosity at $T_g$, B an empirically obtained constant, $T_\infty$
the so-called Vogel temperature and $f_g$ the free liquid volume
fraction.

Besides the above discussed impact of temperature on the viscosity
of a polymer melt, also the molecular weight $M_w$ strongly
influences $\eta$. While $M_w$ increases, the chain mobility
decreases and therefore the relaxation times and thus $\eta$
increase. For sufficiently small $M_w$, the Rouse model predicts a
linear increase of $\eta$ for increasing $M_w$. At this point,
another characteristic number of polymer physics has to be
introduced: the critical chain length for entanglements $M_c$. Above
$M_c$, the polymer chains form chain entanglements exhibiting a
specific mean strand length called \textit{entanglement length}
$M_e$. For PS, $M_c$=35\,kg/mol and $M_e$=17\,kg/mol is found
\cite{Rub03}. According to the reptation model of de Gennes
\cite{deG71}, the viscosity increases stronger in the presence of
chain entanglements and an algebraic behavior of $\eta\propto M_w^3$
is expected. Empirically, the linear regime below $M_c$ is well
reproduced, above $M_c$ an exponent of 3.4 is found experimentally
for different polymers. For a detailed description of chemical and
physical properties of polymers as well as for the Rouse or the
reptation theory we refer to the book \textit{Polymer physics} by
Rubinstein and Colby \cite{Rub03}.

\subsubsection{Properties in confined geometries}
\label{confinement} 
\begin{figure}[t]
\begin{center}
\includegraphics[width=0.5\textwidth]{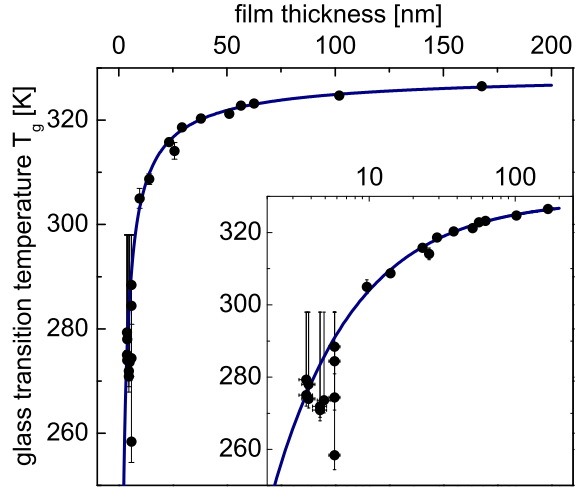}
\end{center}
\caption{Glass transition temperature $T_g$ of polystyrene films of
2\,kg/mol against film thickness (adapted from \cite{Her03}).}
\label{graphtg}
\end{figure}

In contrast to the bulk situation, where volume
properties of the polymers are measured, liquids in confined
geometries such as a thin film often show deviations from the
behavior in the volume due to additional interface effects. One of
these properties is the aforementioned glass transition temperature.
It has been shown in numerous studies, that $T_g$ changes with film
thickness $h$. On the one hand, in case of free-standing or
supported films exhibiting no or repulsive interactions with the
substrate, $T_g(h)$ decreases with decreasing film thickness
\cite{Ked94,Mat00,Dal00,Her01}. On the other hand, Keddie and Jones
have shown that an increase of the glass transition temperature with
decreasing film thickness is possible for attractive interactions
between substrate and polymer film \cite{Ked95}. The influence of
the interfacial energy on the deviation of $T_g$ from its bulk value
has been studied and quantified by Fryer and co-workers for
different polymers \cite{Fry01}. In case of PS on a solid support, a
significant change of $T_g$ is found for films thinner than 100\,nm
(see Fig. \ref{graphtg}) \cite{Her03}. For PS(2k) below 10\,nm for
instance, the glass transition temperature and the viscosity of the
polymer film are affected such that these films are liquid at room
temperature and may dewet spinodally.

Several attempts have been made to explain the change of $T_g$
according to the film thickness. Besides interface-related effects
such as reorientation of polymer chains or accumulation of
chain-ends at the interface, finite-size effects have been
proposed to be responsible. Herminghaus \etal discussed the strain
relaxation behavior of thin viscoelastic polymer films with regard
to surface melting and the shift of the glass transition
temperature \cite{Her03}. Kawana and Jones studied the thermal
expansivity of thin supported polymer films using ellipsometry and
attributed their results concerning $T_g$ to a liquid-like surface
layer \cite{Kaw01}, a result that was also found by other authors
\cite{Her04,See06}. Besides confinement effects on $T_g$, further
interface-related phenomena have been studied: Si \etal have shown
that polymers in thin films are less entangled than bulk polymers
and that the effective entanglement molecular weight $M_e$ is
significantly larger than the bulk value \cite{Si05}.

\subsubsection{Viscosity and viscoelasticity}
\label{viscoelasticity}

One of the major characteristics of a liquid in general or a polymer
in particular is its viscosity $\eta$. Applying shear stress
$\sigma$ to a liquid, it usually reacts with a strain $\gamma$. If
stress and strain rate $\dot\gamma$ are proportional, the fluid is
called \textit{Newtonian}. The constant of proportionality is
identified as the viscosity $\eta$ of the liquid.

\begin{equation}
\label{Newtonian} \sigma=\eta\dot\gamma.
\end{equation}

Liquids such as long-chained polymers show a shear rate dependent
viscosity $\eta(\dot\gamma)$ due to the fact that the liquid
molecules are entangled. If the viscosity increases while shearing
the liquid, we call this behavior \textit{shear thickening}, whereas
in case of lowered viscosity so-called \textit{shear thinning} is
responsible. In contrast to the elastic deformation of a solid, a
deformation of a \textit{viscoelastic} liquid might induce an
additional flow and can relax on a specific time scale $\tau$. On
short time scales ($t<\tau$), the liquid behaves in an elastic, on
long time scales ($t>\tau$) in a viscous manner. Thereby, strain
$\gamma$ is connected to stress $\sigma$ via the elastic modulus $G$
of the liquid:

\begin{equation}
\label{Non-Newtonian} \sigma=G\gamma.
\end{equation}

\begin{figure}[t]
\begin{center}
\includegraphics[width=0.5\textwidth]{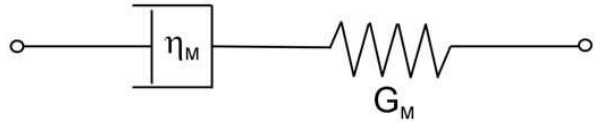}
\end{center}
\caption{Maxwell model represented by a dashpot and a spring in a
serial connection.} \label{graphMaxwell}
\end{figure}

To cover the stress relaxation dynamics of a polymer film, several
modelling attempts have been proposed. Mostly, so-called
\textit{Maxwell} or \textit{Jeffreys} models are applied. The
simplest model is the \textit{Maxwell} model (see Fig.
\ref{graphMaxwell}), which assumes a serial connection of a
perfectly elastic element (represented by a spring) and a
perfectly viscous one (represented by a dashpot). Consequently,
the total shear strain $\gamma$ is given by the sum of the
corresponding shear strains $\gamma_e$ and $\gamma_v$ of both
mechanisms:

\begin{equation}
\label{Maxwell} \gamma=\gamma_e+\gamma_v.
\end{equation}

\noindent With (\ref{Non-Newtonian}) and (\ref{Newtonian}) we get

\begin{equation}
\sigma=G_M \gamma_e=\eta_M \dot \gamma_v
\end{equation}

\noindent since both react on the same shear stress $\sigma$.
Thereby, the ratio of viscosity of the viscous element to the
elastic modulus of the elastic one can be identified with a specific
time scale, the relaxation time $\tau_M$:

\begin{equation}
\tau_M=\frac{\eta_M}{G_M}.
\end{equation}

\noindent The relaxation of stress after a step strain $\gamma$
leads to a time-dependent stress function $\sigma(t)$ for a
viscoelastic liquid. Due to the fact that the total strain $\gamma$
is constant, a first order differential equation for the
time-dependent strain $\gamma_v(t)$ is obtained:

\begin{equation}
\tau_M\dot \gamma_v=\gamma-\gamma_v(t).
\end{equation}

\noindent Solving this differential equation using the initial
condition $\gamma_v(t=0)=0$ gives a simple exponential decay of
$\sigma(t)$ on the time scale of the stress relaxation time $\tau_M$
in the Maxwell model:

\begin{equation}
\gamma_e(t)=\gamma\exp{(-t/\tau_M)}, \\
\sigma(t)=G_M\gamma_e(t)=G_M\gamma\exp{(-t/\tau_M)}.
\end{equation}

\noindent A situation of special interest is the linear response
region: For sufficiently small values of $\gamma$, the
stress/strain-relation (\ref{Non-Newtonian}) is valid and the stress
relaxation modulus $G(t)$ is independent of the strain $\gamma$. In
this regime, a linear superposition of stresses resulting from an
infinite number of strain steps can be used to model a steady simple
shear flow of a viscoelastic liquid. For larger applied shear rates,
linear response and the linear superposition fails. The viscosity is
still defined as the ratio of stress and strain rate, but it has to
be regarded as an apparent viscosity which differs from the above
described ''zero shear rate'' viscosity. Polymers with
\textit{shear-thinning} or \textit{shear-thickening} properties can
be described by the function

\begin{equation}
\sigma\propto \dot \gamma^n,
\end{equation}

\noindent where the exponent $n$ can be extracted from experimental
data. These type of fluids are also called ''power law fluids''.
Moreover, also other nontrivial stress-strain relations can be
considered or alternatively non-linear extensions can be applied to
the linear \textit{Maxwell} models. In case of the linear
\textit{Jeffreys} model, the stress tensor $\sigma_{ij}$ relaxes
according to the following constitutive relaxation equation:

\begin{equation}
\label{Jeffreys} (1 + \lambda_1 \partial_t) \sigma = \eta (1+
\lambda_2 \partial_t) \dot \gamma,
\end{equation}

\noindent where the strain rate is given by the gradient of the
velocity field $\dot\gamma_{ij}=\partial_ju_i+\partial_iu_j$.
Hereby, $\lambda_1$ governs the relaxation of stress, whereas
$\lambda_2$ ($\lambda_2<\lambda_1$) describes the relaxation of the
strain rate, respectively. This model accounts for the viscous and
the elastic properties of a fluid and was used by Blossey, Rauscher,
Wagner and M\"unch as basis for the development of a thin-film
equation that incorporates viscoelastic effects
\cite{Rau05,Blo06,Mue06}. For a more elaborate description of
non-Newtonian flows we refer e.g. to the correspondent work of Te
Nijenhuis \etal \cite{Nij07}.

\subsubsection{Reynolds and Weissenberg number}
The flow of a liquid can be characterized by specific numbers. One
of these numbers is the so-called Reynolds number \textit{Re}, which
describes the ratio of inertia effects to viscous flow
contributions. In case of thin liquid films, \textit{Re} can be
written as

\begin{equation}
\label{Reynolds} Re=\frac{\rho u h}{\eta},
\end{equation}

\noindent where $\rho$ denotes the density of the liquid, $u$
describes the flow velocity and $h$ stands for the film thickness
\cite{Mue05}. For thin dewetting polymer films, the Reynolds number
is very small, i.e. $Re\ll1$, and a low-\textit{Re} lubrication
theory can be applied. To quantify and to judge the occurrence of
viscoelastic effects versus pure viscous flow, the so-called
Weissenberg number $Wi$ has been introduced as

\begin{equation}
\label{Weissenberg} Wi=\tau\dot\gamma.
\end{equation}

\noindent Thereby, $\tau$ denotes the relaxation time and
$\dot\gamma$ the strain rate as introduced in the previous section.
If $Wi\ll1$, an impact of viscoelasticity on flow dynamics can be
neglected and viscous flow dominates.

\subsection{Navier-Stokes equations}

The Navier-Stokes equations for a Newtonian liquid mark the
starting point for the discussion of fluid dynamics in confined
geometries. According to conservation of mass, the equation of
continuity can be formulated as

\begin{equation}
\label{Kontinuitätsgleichung}
\partial_t\rho+\nabla\cdot(\rho \textbf{u})=0,
\end{equation}

\noindent where $\textbf{u}=(u_x,u_y,u_z)$ is the velocity field of
the fluid. For an incompressible liquid, which implies a temporally
and spatially constant liquid density $\rho$,
(\ref{Kontinuitätsgleichung}) can be simplified to

\begin{equation}
\label{Kontinuitätsgleichung2} \nabla \textbf{u}=0.
\end{equation}

\noindent With the conservation of momentum, the Navier-Stokes
equations for an incompressible liquid can be written as

\begin{equation}
\label{NS} \rho(\partial_t+ \textbf{u}\cdot\nabla)\textbf{u}=-\nabla
p + \eta\triangle \textbf{u} + \textbf{f},
\end{equation}

\noindent with the pressure gradient $\nabla p$ and the volume force
$\textbf{f}$ of external fields acting as driving forces for the
liquid flow. We already stated that for small Reynolds numbers, i.e.
$Re\ll1$, the terms on the left hand side of (\ref{NS}) can be
neglected as compared to terms describing the pressure gradient,
external volume forces and viscous flow. By that, we can simplify
(\ref{NS}) to the so-called Stokes equation

\begin{equation}
\label{Stokes} 0=-\nabla p + \eta\triangle \textbf{u} + \textbf{f}.
\end{equation}

\noindent In section \ref{TFsection}, we will demonstrate how these
basic laws of bulk fluid dynamics can be applied to the flow
geometry of a thin film supported by a solid substrate.

\subsection{Free interface boundary condition}

At the free interface of a supported liquid film, i.e. at the
liquid/gas or usually the liquid/air interface, no shear forces can
be transferred to the gas phase due to the negligible viscosity of
the gas. In general, the stress tensor $\sigma_{ij}^\ast$ is given
by the stress tension $\sigma_{ij}$, see (\ref{Newtonian}), and the
pressure $p$:

\begin{equation}
\label{straintensor}
\sigma_{ij}^\ast=\sigma_{ij}+p\delta_{ij}=\eta(\partial_ju_i+\partial_iu_j)+p\delta_{ij}.
\end{equation}

\noindent The tangential \textbf{t} and normal \textbf{n}
(perpendicular to the interface) components of the stress tensor
are:

\begin{equation}
\label{compST} (\sigma^\ast\cdot\textbf{n})\cdot\textbf{t}=0 \\
(\sigma^\ast\cdot\textbf{n})\cdot\textbf{n}=\gamma_{lv}\kappa,
\end{equation}

\noindent where $\kappa$ denotes the mean curvature and
$\gamma_{lv}$ the interfacial tension (i.e. the surface tension of
the liquid) of the liquid/vapor interface. If the liquid is at rest,
i.e. the stationary case $\textbf{u}=0$, the latter boundary
condition gives the equation for the Laplace pressure $p_L$:

\begin{equation}
\label{Laplace}
p_L=\gamma_{lv}\kappa=\gamma_{lv}(\frac{1}{R_1}+\frac{1}{R_2}).
\end{equation}

\noindent $R_1$ and $R_2$ are the principal radii of curvature of
the free liquid/gas boundary; the appropriate signs of the radii are
chosen according to the condition that convex boundaries give
positive signs. Such convex liquid/gas boundaries lead to an
additional pressure within the liquid due to its surface tension. In
the next section, the solid/liquid boundary condition will be
discussed, which yields a treatment of slip effects.

\subsection{Slip/no-slip boundary condition}
\label{Navier}

\subsubsection{Navier slip boundary condition}

In contrast to fluid dynamics in a bulk volume, where the assumption
that the tangential velocity \textbf{u}$_{||}$ at the solid/liquid
interface vanishes (\textit{no-slip boundary condition}), confined
geometries require a more detailed investigation as slippage becomes
important. In 1823, Navier \cite{Nav23} introduced a linear boundary
condition: The tangential velocity \textbf{u}$_{||}$ is proportional
to the normal component of the strain rate tensor; the constant of
proportionality is described as the so-called \textit{slip length}
$b$:

\begin{figure}[t]
\begin{center}
\includegraphics[width=0.9\textwidth]{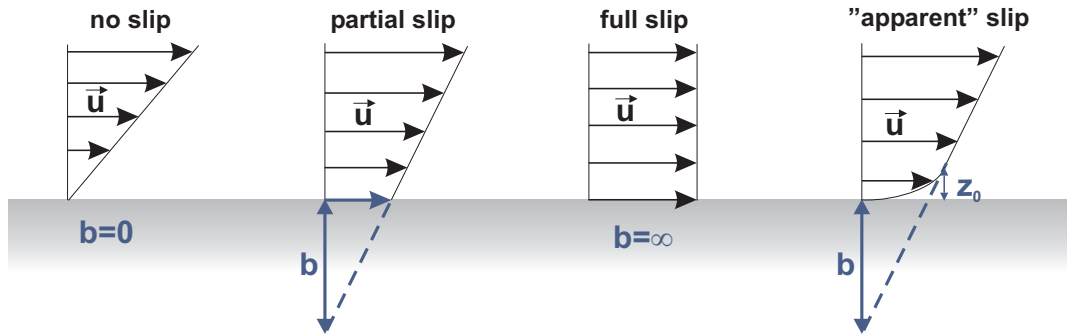}
\end{center}
\caption{Different velocity profiles in the vicinity of the
solid/liquid interface and illustration of the slip (extrapolation)
length $b$. The situation of so-called ''apparent'' slip is
illustrated on the right: According to a thin liquid layer of
thickness $z_0$ that obtains a significantly reduced viscosity, the
slip velocity $u_x|_{z=0}$ is zero, but a substantial slip length is
measured.} \label{graphSlip}
\end{figure}

\begin{equation}
\label{sliplength} \textbf{u}_{||}=b\,\textbf{n}\cdot\dot\gamma
\end{equation}

\noindent In case of simple shear flow in \textit{x}-direction, the
definition of the slip length can be alternatively written as

\begin{equation}
\label{sliplength2}
b=\frac{u_x}{\partial_zu_x}|_{z=0}=\frac{u_x\eta}{\sigma}=\frac{\eta}{\xi},
\end{equation}

\noindent where $\xi=\sigma/u_x$ denotes the friction coefficient at
the solid/liquid interface. The \textit{xy}-plane thereby represents
the substrate surface. According to these definitions, the slip
length can be illustrated as the extrapolation length of the
velocity profile ''inside'' the substrate, cf. Fig. \ref{graphSlip}.
Moreover, both limiting cases are included within this description:
For $b=0$, we obtain the \textit{no-slip} situation, whereas
$b=\infty$ characterizes a \textit{full-slip} situation. The latter
case corresponds to ''plug-flow'', where the liquid behaves like a
solid that slips over the support.

\subsubsection{How to measure the slip length?}

In recent years, numerous experimental studies were published
using diverse methods to probe the slip length at the boundary of
different simple or complex liquids and solid supports. For
details concerning these experimental methods we refer to the
review articles from Lauga \etal \cite{Lau07}, Neto \etal
\cite{Net05} and Bocquet and Barrat \cite{Boc07} (and references
therein). To probe the boundary condition, scientists performed
either drainage experiments or direct measurements of the local
velocity profile using e.g. tracer particles.

In case of drainage experiments, the liquid is squeezed between two
objects, e.g. a flat surface and a colloidal probe at the tip of an
AFM cantilever, and the corresponding force for dragging the probe
is measured (\textit{colloidal probe AFM}). Alternatively, in an
\textit{surface force apparatus (SFA)}, two cylinders arranged
perpendicular to each other are brought in closer contact and
force/distance measurements are performed to infer the slip length.

The use of tracer particles as a probe of the local flow profile
might bring some disadvantages. The chemistry of these particles is
usually different from the liquid molecules and their influence on
the results might not be negligible. A similar method is called
\textit{fluorescence recovery after photo bleaching}. Thereby, a
distinct part of a fluorescent liquid is bleached by a laser pulse
and the flow of non-bleached liquid into that part is measured. The
disadvantage of this method is that diffusion might be a further
parameter that is hard to control. Recently, Joly \etal showed that
also thermal motion of confined colloidal tracers in the vicinity of
the solid/liquid interface can be used as a probe of slippage
without relying on external driving forces \cite{Jol06}.

\subsubsection{Which parameters influence slippage?}

Of course, many interesting aspects in the field of micro- and
nanofluidics are related to intrinsic parameters that govern
slippage of liquid molecules at the solid/liquid interface. For
simple liquids on smooth surfaces, the contact angle is one of the
main parameters influencing slippage
\cite{Bar991,Pit00,Cot02,Leg03}. This originates from the effect of
molecular interactions between liquid molecules and the solid
surface: If the molecular attraction of liquid molecules and surface
decreases (and thereby the contact angle increases), slippage is
enlarged. Further studies aim to quantify the impact of roughness
\cite{Pit00,Zhu02,Sch06,Kun07} or topographic structure
\cite{Cot03,Pri06,Jos06,Ybe07,Ste07} of the surface on slippage. For
different roughness length scales, a suppression (see e.g.
\cite{Pit00,Zhu02}) or an amplification (see e.g. \cite{Cot03}) of
slippage can be observed. Moreover, the shape of molecular liquids
itself has been experimentally shown to impact the boundary
condition. Schmatko \etal found significantly larger slip lengths
for elongated linear compared to branched molecules \cite{Sch05}.
This might be associated with molecular ordering effects
\cite{Pri05} and the formation of layers of the fluid in case of the
capability of these liquids to align in the vicinity of the
interface \cite{Hei07}. Cho \etal identified the dipole moment of
Newtonian liquids at hydrophobic surfaces as a crucial parameter for
slip \cite{Cho04}. De Gennes proposed a thin gas layer at the
interface of solid surface and liquid as a possible source of large
slip lengths \cite{deG02}. Recently, MD studies for water on
hydrophobic surfaces by Huang \etal revealed a dependence of
slippage on the amount of water depletion at the surface and a
strong increase of slip with increasing contact angle \cite{Hua08}.
Such depletion layers for water in the vicinity of smooth
hydrophobic surfaces have also been experimentally observed using
scattering techniques \cite{Ste03,Dos05,Mez06,Mac07}. Contamination
by nanoscale air bubbles (so-called nanobubbles) and its influence
on slippage has been controversially discussed in literature (see
e.g. \cite{Tyr01,Tre04,Poy06,Hen09}). In case of more complex
liquids such as polymer melts further concepts come into play. They
will be illustrated in section \ref{shear-dependent-slippage}.

\subsection{Thin-film equation for Newtonian liquids}
\label{TFsection}

\subsubsection{Derivation} Confining the flow of a liquid to the
geometry to the one of a thin film, we can assume that the velocity
contribution perpendicular to the substrate is much smaller than the
parallel one. Furthermore, the lateral length scale of film
thickness variations is much smaller than the film thickness itself.
On the basis of these assumptions, Oron \etal \cite{Oro97} developed
a thin-film equation from the rather complex equations of motion,
(\ref{Kontinuitätsgleichung}) and (\ref{NS}). In case of film
thicknesses smaller than the capillary length
$l_c=\sqrt{\gamma_{lv}/\rho g}$, (which is typically in the order of
magnitude of 1\,mm) (\ref{Stokes}) can be written as

\begin{equation}
\label{TF1} 0=-\nabla(p+\phi^\prime(h)) + \eta\triangle \textbf{u}
\end{equation}

\noindent Additional external fields such as gravitation can be
neglected, but a secondary contribution $\phi^\prime(h)$, the
\textit{disjoining pressure}, has been added to the capillary
pressure $p$. The disjoining pressure originates from molecular
interactions of the fluid molecules with the substrate. The
effective interface potential $\phi(h)$ summarizes the
inter-molecular interactions and describes the energy that is
required to bring two interfaces from infinity to the finite
distance $h$. As already discussed in the introductory part, the
stability of a thin liquid film is also governed by $\phi(h)$. For a
further description of thin film stability, we refer to \cite{Jac08}
and the references therein.

\begin{figure}[t]
\begin{center}
\includegraphics[width=0.5\textwidth]{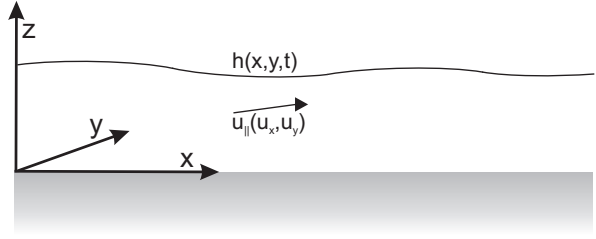}
\end{center}
\caption{Illustration of the nomenclature of the thin film length
scales ($x$ and $y$ are parallel to the substrate) and the velocity
contribution \textbf{u}$_{||}=(u_x, u_y)$.} \label{graphThinfilm}
\end{figure}

The derivation of a thin-film equation for Newtonian liquids starts
with the kinematic condition

\begin{equation}
\label{KK}
\partial_th=-\nabla_{||}\int_0^{h}\textbf{u}_{||} dz,
\end{equation}

\noindent i.e. the coupling of the time derivative of $h(x,y,t)$ to
the flow field, where the index $||$ in general denotes the
components parallel to the substrate ($\nabla_{||}=(\partial_x,
\partial_y)$ and \textbf{u}$_{||}=(u_x, u_y)$) as illustrated in Fig. \ref{graphThinfilm}.

For thin liquid films, film thickness variations on lateral scale
$\mathbb{L}$ are much larger than the length scale of the film
thickness $\mathbb{H}$. Introducing the parameter
$\epsilon=\mathbb{H}/\mathbb{L} \ll1$ yields the so-called
\textit{lubrication approximation} and is used in the following to
re-scale the variables to dimensionless values. In a first
approximation, linearized equations are obtained while neglecting
all terms of the order $O(\epsilon^2)$. For reasons of simplicity
and due to translational invariance in the surface plane, a
one-dimensional geometry is used:

\begin{equation}
\label{linearized}
\partial_x(p+\phi^\prime)=\partial_{z}^2u_x, \\
\partial_z(p+\phi^\prime)=0, \\
\partial_xu_x+\partial_zu_z=0.
\end{equation}

\noindent While the substrate is supposed to be impenetrable for the
liquid, i.e. $u_z=0$ for $z=0$, friction at the interface implies a
velocity gradient $\partial_zu_x=u_x/b$ for $z=0$. Moreover, the
tangential and normal boundary condition at the free interface, i.e.
$z=h(x)$, can be simplified in the following manner:

\begin{equation}
\partial_zu_x=0, \\
p+\partial_x^2h=0.
\end{equation}

From (\ref{linearized}) and the boundary conditions, the velocity
profile $u_x(z)$ can be obtained. Using the kinematic condition
(\ref{KK}), the equation of motion for thin films in three
dimensions is derived:

\begin{equation}
\label{TFE}
\partial_t h=-\nabla[m(h)\nabla(\gamma_{lv}\triangle
h-\phi^\prime(h))],
\end{equation}

\noindent where $m(h)$ denotes the mobility given by

\begin{equation}
\label{mobility} m(h)=\frac{1}{3\eta}(h^3+3bh^2).
\end{equation}

\subsubsection{Lubrication models including slippage}
As discussed in the previous section, the derivation of the
thin-film equation is based on the so-called lubrication
approximation and the re-scaling of relevant values in $\epsilon$.
As a consequence, the slip length $b$ is supposed to obtain values
smaller than the film thickness $h$, i.e. $b\ll h$. To extend this
so-called \textit{weak-slip} situation with regard to larger slip
$b\gg h$, M\"unch \etal \cite{Mue05} and Kargupta \etal \cite{Kar04}
developed independently so-called \textit{strong-slip} models.
Thereby, the slip length is defined as $b=\beta/\epsilon^2$. The
corresponding equation of motion together with the kinematic
condition in one dimension for a Newtonian thin liquid film read as:

\begin{equation}
\label{strongslip} \eqalign{u=\frac{2b}{\eta}\partial_x(2\eta
h\partial_x u)+\frac{bh}{\eta}\partial_x(\gamma_{lv}\partial_x^2
h-\phi^\prime(h)) \\
\partial_t h=-\partial_x(hu).}
\end{equation}

In fact, a family of lubrications models, cf. Tab. \ref{table1},
accounting for different slip situations have been derived. In the
limit $b\rightarrow 0$, i.e. the \textit{no-slip} situation, the
mobility is given by $m(h)=h^3/3\eta$. If the slip length is in the
range of the film thickness $b\sim h$, the mobility in the
corresponding \textit{intermediate-slip} model is $m(h)=bh^2/\eta$.
Recently, Fetzer \etal \cite{Fet07} derived a more generalized model
based on the full Stokes equations, developed up to third order of a
Taylor expansion. The authors were able to show that this model is
in good agreement with numerical simulations of the full
hydrodynamic equations and is not restricted to a certain slip
regime as the aforementioned lubrication models.

\begin{table}
\caption{\label{table1}Summary of lubrication models for Newtonian
flow and different slip situations.}
\begin{indented}
\item[]\begin{tabular}{@{}lllll}
\br
    model & validity & equation & limiting cases & ref.\\
\mr
    \multirow{2}{*}{weak-slip} & \multirow{2}{*}{$b\ll h$} &
    \multirow{2}{*}{(\ref{TFE}),
    (\ref{mobility})} &  $b\rightarrow 0$ (no-slip) &
    \multirow{2}{*}{\cite{Oro97}}
    \\ & & & $b\rightarrow \infty$ (intermediate-slip) & \\
\mr
    \multirow{2}{*}{strong-slip} & \multirow{2}{*}{$b\gg h$} & \multirow{2}{*}{(\ref{strongslip})} &
    $\beta\rightarrow 0$ (intermediate-slip) & \multirow{2}{*}{\cite{Kar04,Mue05}}
    \\ & & & $\beta\rightarrow \infty$ (''free''-slip) & \\
\br
\end{tabular}
\end{indented}
\end{table}

\subsubsection{Lubrication models including viscoelasticity}
In the meanwhile, the derivation of a thin film equation for the
weak-slip case including linear viscoelastic effects of Jeffreys
type (such as described by equation (\ref{Jeffreys}) in section
\ref{viscoelasticity}) has been achieved (see \cite{Rau05}). To
cover relaxation dynamics of the stress tensor $\sigma$, an
additional term $\nabla \cdot \sigma$ on the right hand side of
(\ref{TF1}) has to be included to the aforementioned model for
Newtonian liquids. Furthermore, the treatment of linear viscoelastic
effects was also achieved for the strong-slip situation by Blossey
\etal \cite{Blo06}. To summarize these extensions, the essential
result is the fact that linear viscoelastic effects are absent in
the weak-slip case and the Newtonian thin-film model is still valid.
The strong-slip situation, however, is more complicated. Slippage
and viscoelasticity are combined and strongly affect the
corresponding equations. In the meanwhile, the authors were able to
fully incorporate the non-linearities of the co-rotational Jeffreys
model for viscoelastic relaxation into their thin-film model
\cite{Mue06}.

\noindent The extensions of the aforementioned thin-film models for
different slip conditions with or without the presence of
viscoelastic relaxation (Newtonian and non-Newtonian models) affect
on the one hand the rupture conditions, but also on the other hand
the shape of a liquid ridge. These two phenomena will be discussed
in the next two subsections. A elaborate description of these
theoretical aspects can be found in a recent review article by
Blossey \cite{Blo08}.

\subsubsection{Application I - Spinodal dewetting}
\label{spinodaldewetting} 
\begin{figure}[t]
\begin{center}
\includegraphics[width=0.6\textwidth]{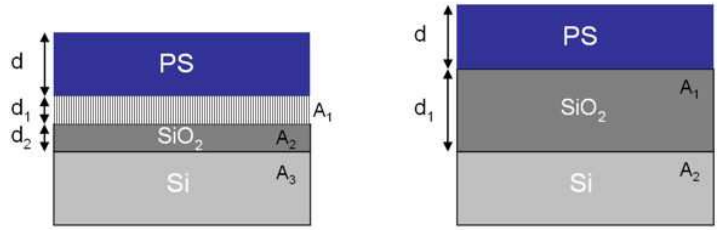}
\end{center}
\caption{Two examples for polystyrene films prepared on multi-layer
substrates. Left: Silicon wafer with native oxide layer covered with
a hydrophobic layer (e.g. a self-assembled monolayer (SAM)). Right:
Silicon wafer with an increased (compared to a native silicon oxide)
oxide layer thickness.} \label{graphLayers}
\end{figure}

One of the main applications of the
theoretical thin-film models is the dewetting of thin polymer films.
As introduced in section \ref{Intro} and illustrated by Fig.
\ref{graphphiskizze}, the stability of a thin liquid film is
governed by the effective interface potential. Basically, long-range
attractive van der Waals forces add to short-range repulsive forces.
Due to the planar geometry of two interfaces of distance $d$, the
van der Waals contribution to the potential is $\phi(d)_{vdW}\propto
-A/d^2$, where $A$ is the Hamaker constant. For the description of
the explicit calculation of Hamaker constants from the dielectric
functions of the involved materials we refer to \cite{Jac08} and to
the book by Israelachvili \cite{Isr92}. Experimental systems often
exhibit multi-layer situations, cf. Fig. \ref{graphLayers}. A
hydrophobic film and/or an oxide layer of distinct thicknesses $d_i$
exhibiting Hamaker constants $A_i$ require a superposition of
contributions to the potential:

\begin{equation}
\label{vdW} \phi(d)_{vdw}=-\frac{A_1}{12\pi
d^2}-\frac{A_2-A_1}{12\pi(d+d_1)^2}-\frac{A_3-A_2}{12\pi(d+d_1+d_2)^2}
\end{equation}

\begin{figure}[t]
\begin{center}
\includegraphics[width=0.6\textwidth]{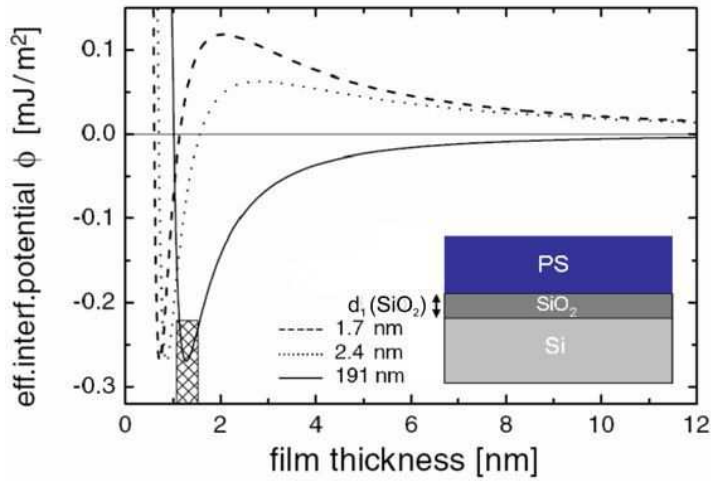}
\end{center}
\caption{Effective interface potential $\phi(h)$, gained from
experimental data, plotted against the thickness $h$ of a thin PS
film prepared on silicon wafers with different oxide layer
thicknesses $d_1$ (adapted from \cite{See012}). The cross-hatched
rectangle marks the error for $h_{eq}$ and the depth of the global
minimum.} \label{graphphi}
\end{figure}

\noindent Consequently, the shape of the effective interface
potential (and thereby also the thin film stability) is governed by
the set of Hamaker constants $A_i$ and film thicknesses $d_i$. E.g.
for a thin PS film of thickness $h$ on a Si substrate with a native
oxide layer of 2.4\,nm, $\phi(h)$ shows a global minimum at an
equilibrium film thickness $h_{eq}$ (c.f. Fig. \ref{graphphi}).
Moreover, a local maximum at $h>h_{eq}$ is found.

If the PS film is sufficiently thin ($\phi^{\prime\prime}(h)<0$) ,
it may become unstable due to the amplification of thermally induced
capillary waves. To track the evolution of small fluctuations of the
film thickness $h$, i.e. $f(\textbf{x},t)=h(\textbf{x},t)-h$ with
$f(\textbf{x},t)\ll h$, a Fourier transform of the linearized thin
film equation (\ref{TFE}) has to be performed. The amplitudes of the
capillary waves grow exponentially in time. Their growth rate
$\alpha$ can be calculated as a function of the wavenumber $q$ and
depends on the sign of the local curvature of the interface
potential. If the second derivative of the effective interface
potential $\phi^{\prime\prime}$ at the film thickness $h$ is
positive, $\alpha$ is negative for all $q$ and the amplitudes of the
capillary waves are damped. If $\phi^{\prime\prime}<0$, the growth
rate $\alpha$ is positive for a certain range of wavenumbers up to a
critical wavenumber $q_c$ and capillary waves are amplified. The
wavenumber that corresponds to the maximum value of $\alpha$ and
therefore exhibits the fastest amplification, is called preferred
wavenumber $q_0$ and is connected to the preferred wavelength
$\lambda_0=2\pi/q_0$. The latter is also called spinodal wavelength
and can be written as

\begin{equation}
\label{PWL} \lambda_0=\sqrt{\frac{8\pi^2
\gamma_{lv}}{-\phi^{\prime\prime}(h)}}.
\end{equation}

\begin{figure}[t]
\begin{center}
\includegraphics[width=0.95\textwidth]{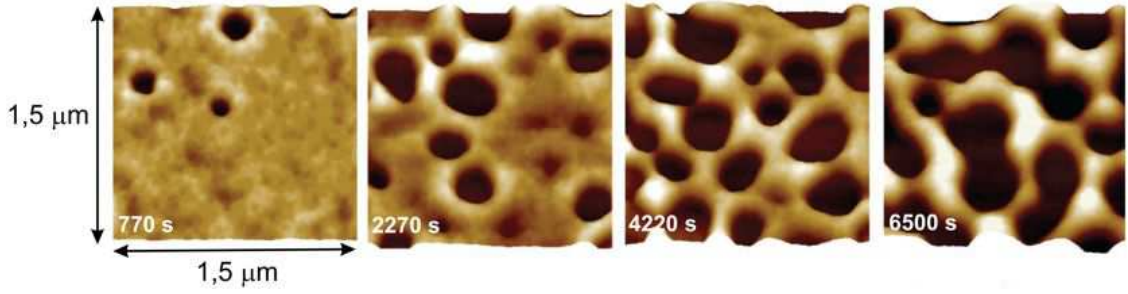}
\caption{\textit{In situ} (at elevated temperature
$T=53\,^{\circ}$C) atomic force microscopy (AFM) images
(corresponding annealing times given in pictures) of a spinodally
dewetting 3.9(2)\,nm PS(2\,kg/mol) film on a Si wafer with a thick
(191\,nm) oxide layer (adapted from
\cite{Bec03}).}\label{graphBecker}
\end{center}
\end{figure}

\noindent The spinodal dewetting process can be monitored e.g. by
atomic force microscopy (AFM) as shown in Fig. \ref{graphBecker}. By
measuring $\lambda_0$ as a function of film thickness,
$\phi^{\prime\prime}(h)$ can be inferred and conclusions can be
drawn with regard to the effective interface potential $\phi(h)$
\cite{See012}. For further details concerning stability of thin
films we refer to \cite{Jac08} and references therein.

Using the strong-slip model while taking slip into account
(\ref{strongslip}), Rauscher and coworkers could theoretically show
that slippage is supposed to influence the capillary wave spectrum
due to a different mobility at the solid/liquid interface
\cite{Rau08}. The position of the maximum $q_0$ shifts to smaller
wavenumbers and larger wavelengths for increasing slip length. As
for today, to the best of our knowledge experimental studies
concerning the impact of slippage on the spinodal wavelength are not
available.

\noindent The description of the influence of thermal noise on the
temporal and spatial dynamics of spinodally dewetting thin polymer
films has been recently achieved by Fetzer and coworkers
\cite{Fet072}. A stochastic Navier-Stokes equation with an
additional random stress fluctuation tensor that accounts for
thermal molecular motion is utilized to model the flow while
assuming a no-slip boundary condition at the solid/liquid interface.
The stochastic model matches the experimentally observed spectrum of
capillary waves and thermal fluctuations cause the coarsening of
typical length scales.

\subsubsection{Application II - Shape of a dewetting rim}
\label{shapeofadewettingrim} 
\begin{figure}[t]
\begin{center}
\includegraphics[width=0.85\textwidth]{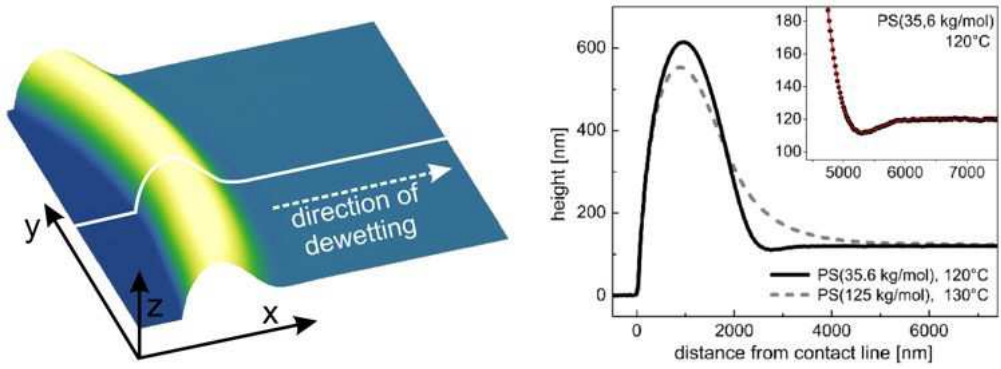}
\end{center}
\caption{Left: AFM image (scan size 10\,$\mu$m) of a liquid rim
formed during hole growth in a PS film dewetting from a hydrophobic
substrate (Si wafer covered with a 21\,nm hydrophobic
Teflon$^{\circledR}$ coating (AF\,1600)). Right: Different rim
morphologies for PS on AF\,1600 (AFM cross-sections): Profile
exhibiting a trough (depicted in the inset) for PS(35.6\, kg/mol) at
120$\,^{\circ}$C and a monotonically decaying rim shape for
PS(125\,kg/mol) at 130$\,^{\circ}$C.} \label{graphRims}
\end{figure}

Besides the implications on spinodal
dewetting, the one-dimensional thin-film model has been successfully
applied to the shape of the rim along the perimeter of e.g.
nucleated holes. Experimentally, researchers have studied and
observed different types of rim profiles \cite{See01,Rei01,Dam03}.
As shown in Fig. \ref{graphRims}, profiles either decay
monotonically into the undisturbed film or they show a more
symmetrical profile exhibiting a trough (termed ''oscillatory
shape''). If the depth of this through is in the range of the film
thickness, a ring of so-called satellite holes can be generated
\cite{Net03-1,Net03-2}.

The shape of a dewetting rim can be understood by the aforementioned
thin film theory for Newtonian liquids: Introducing a small
perturbation $\delta h(x,t) \ll h$ of the film thickness $h(x,t)$
and small velocities $u(x,t)$ leads to linearized thin-film
equations that describe the temporal and spatial evolution of
$\delta h$. Thereby, the disjoining pressure $\phi^{\prime}(h)$ can
be neglected due to the fact that films thicker than 10\,nm are
considered. To obtain stationary solutions of the linearized
equations, a frame that is co-moving with the rim
$\zeta(x,t)=x-s(t)$ is introduced. Thereby, $s(t)$ denotes the
position of the three-phase contact line; $\dot s$ stands for the
dewetting velocity $V$ as described in the next sections. Fetzer
\etal used a normal modes ansatz $\delta h(\zeta)=\delta h_0
\exp{k\zeta}$ and $u(\zeta)=u_0 \exp{k\zeta}$ in the linear
stability analysis, which leads to a characteristic polynomial of
third order. Depending on the ratio of slip length to film thickness
$b/h$ and on the capillary number $Ca=\eta\dot s/\gamma_{lv}$, the
parameter $k$ obtains complex or real solutions. Fetzer and
coworkers successfully identified the morphological transition from
oscillatory to monotonically decaying rims and were able to extract
slip lengths and capillary numbers from diverse experiments on
dewetting surfaces \cite{Fet05,Fet06,Fet07,Bae09}.

\section{Flow dynamics of thin polymer films - Experimental studies and theoretical models}

\begin{figure}[t]
\begin{center}
\includegraphics[width=0.95\textwidth]{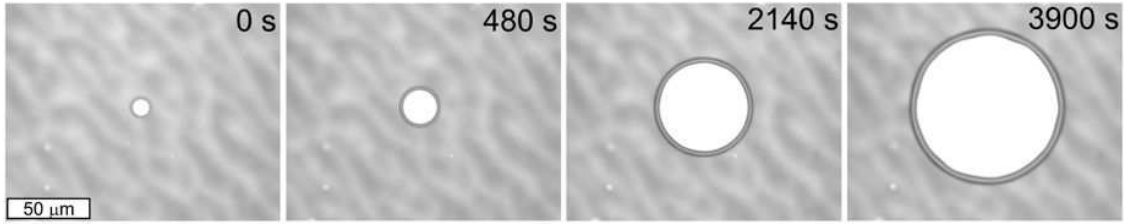}
\end{center}
\caption{Temporal series of optical micrographs showing the growth
of a hole in a 120\,nm PS(13.7\,kg/mol) film at $T=120\,^{\circ}$C
prepared on a Si wafer covered with a 21\,nm hydrophobic
Teflon$^{\circledR}$ coating (AF\,1600). In the last stage of hole
growth (right image) perturbations of the three-phase contact line
(according to the liquid rim instability) become visible.}
\label{graphHoles}
\end{figure}

One of the main aspects of experimental studies concerning the flow
dynamics of thin polymer films is to obtain a comprehensive view on
the molecular mechanisms of slippage and on the responsible
parameters. Although these insights are rather indirect, several
models have been proposed to explain diverse experimental results.
In this section, we will focus on these studies, with special regard
to the proposed mechanisms of slippage at the solid/polymer
interface. In general, we have to distinguish two different
dewetting geometries: growth of holes (cf. Fig. \ref{graphHoles})
and receding straight fronts. On the one hand driving forces and on
the other hand dissipation mechanisms have to be considered.

\subsection{Dewetting dynamics - Driving forces} According to the
description of the effective interface potential in section
\ref{spinodaldewetting}, a global minimum of $\phi(h)$ occurs at
$h_{eq}$ in case of unstable or metastable films. This means that
the film will thin until a thickness of $h_{eq}$ is reached. In
other words, a thin wetting layer remains on top of the substrate,
if $h_{eq}>0$ and if $h_{eq}$ has a size that is not below the size
of the molecules \footnote{Note that the concept of the effective
interface potential is a continuum approach. It may fail if
molecular sized phenomena are to be captured, such as extremely thin
films close to the size of the molecules.}. After dewetting has
taken place (the dynamics of which is not covered by $\phi(h)$ but
depends on viscosity, viscoelasticity, contact angle and the
solid/liquid boundary condition), single droplets remain on top of
the wetting layer. The droplets - in equilibrium - exhibit the
Young's contact angle $\theta_Y$. Parallel to the Young's equation
that characterizes the contact angle via the involved surface
tensions, another characteristic parameter can be defined to specify
the wettability of a surface by a liquid. This is the so-called
\textit{spreading parameter} $S$ given by

\begin{equation}
\label{Spreading}
S=\gamma_{sv}-(\gamma_{sl}+\gamma_{lv}+\phi(h_{eq})),
\end{equation}

\noindent where $\gamma_{ij}$ denotes the interface tension of the
solid (\textit{s}), liquid (\textit{l}) and the vapor (\textit{v})
phase. $S$ describes the energetic difference per unit area between
a dry surface and a surface that is covered by a liquid layer.
Consequently, if $S<0$ the system can gain energy by reducing the
contact area with the substrate. Thereby, the liquid forms a dynamic
contact angle $\theta_d=\theta_Y/\sqrt{2}$ at the three-phase
contact line \cite{Red91}.

The link between the contact angle in equilibrium and the depth of
the minimum of the effective interface potential becomes obvious, if
the equation for the Young's contact angle is inserted in
(\ref{Spreading}):

\begin{equation}
\label{Spreading2} \phi(h_{eq})=\gamma_{lv}(\cos\theta_Y-1)=S
\end{equation}

Driving forces for wetting and dewetting are capillary forces (see
\cite{deG85}). Thereby, the capillary force per unit length of the
contact line is in general given by

\begin{equation}
\label{capillary}
\frac{F_c}{l}=\gamma_{lv}(\cos{\theta_Y}-\cos{\theta_d}).
\end{equation}

\noindent As long as $\theta_d<\theta_Y$ remains valid, the
capillary force is negative and the three-phase contact line will
recede; the liquid film dewets. Dewetting ends as soon as droplets
exhibit their Young's contact angle $\theta_Y$ on the surface.
During dewetting, a rim which is formed from accumulated material
and retracts from the substrate while further growing. De Gennes
developed a theoretical description of the resulting driving force
for dewetting \cite{deG85}:

On the ''dry'' side of the rim, which denotes the side where the
three-phase contact line is located, a negative capillary force
pulls at the contact line:

\begin{equation}
\label{capillarydry} \frac{F_c}{l}=S+\gamma_{lv}(1-cos{\theta_d}).
\end{equation}

\noindent On the other side, where the rim decays into the liquid
film (the ''wet'' side of the rim), a positive capillary force
occurs:

\begin{equation}
\label{capillarywet} \frac{F_c}{l}=\gamma_{lv}(1-cos{\theta_d}).
\end{equation}

\noindent We have to remark that in experimental systems, the
Laplace pressure of course avoids edges and therefore forces the dry
side of the rim to decay smoothly into the film. Summing up both
forces per unit length gives the effective driving force for
retracting contact lines in dewetting systems:

\begin{equation}
\label{driving}
\frac{F_c}{l}=|S|=\gamma_{lv}(1-\cos\theta_Y)\simeq
\frac{1}{2}\gamma_{lv}\theta_Y.
\end{equation}

\noindent Thereby, it is assumed that the system is in a
quasi-stationary state, which means that changes in the shape of
the rim occur much slower that the dewetting velocity. All in all,
this result implies that the driving force for dewetting is given
by the absolute value of the spreading parameter, which solely
depends on the surface tension $\gamma_{lv}$ of the liquid (a
property which is purely inherent to the liquid) and the Young's
contact angle of the liquid on the surface. The same conclusion is
drawn if considering energies instead of driving forces. The depth
of the minimum of the effective interface potential $\phi(h)$ is
equal to $S$ and gives the energy per unit area that is set free
during dewetting \cite{Fru38}. Thereby, $S$ also stands for the
force per unit length of the contact line (see (\ref{driving})).

\subsection{Dewetting dynamics - Dissipation mechanisms}

In the previous section, we focused on the driving forces for
dewetting. The experimentally observed dynamics represents a force
balance of driving forces and friction forces. The resulting
dewetting velocity $V$ is connected to the spreading parameter $S$
and the occurring energy dissipation mechanisms ($F_i$ gives the
friction force per unit length of the contact line) and their
corresponding velocity contributions $v_i$ via the power balance

\begin{equation}
\label{forcebalance} |S|V=\sum_i{F_iv_i}.
\end{equation}

\noindent In case of dissipation due to viscous friction within the
liquid ($F_v$) and dissipation due to friction of liquid molecules
at the solid/liquid interface ($F_s$), i.e. slippage with a finite
velocity $v_s=u|_{z=0}$, we end up with the following balance (the
index $v$ stands for 'viscous', $s$ denotes 'slip'):

\begin{equation}
\label{forcebalance2} |S|V=F_vv_v+F_sv_s.
\end{equation}

\subsubsection{Viscous friction}
\label{viscous friction} According to the work of Brochard-Wyart
\etal \cite{Bro94}, the no-slip boundary condition for fluid flow
implies a friction force that is proportional to the liquid
viscosity and the dewetting velocity $v$ and means that the
dissipation is solely due to viscous friction within the liquid. The
largest strain rates occur in the direct vicinity of the three-phase
contact line. The dissipation is therefore mainly independent from
the size of the dewetting rim. However, the flow geometry (given by
the dynamic contact angle $\theta_d$) at the contact line influences
viscous dissipation. In case of pure viscous flow $V=v_v$, the
following expression for the velocity contribution is obtained for
viscous flow \cite{Bro94}:

\begin{equation}
\label{viscousflow} v_v=C_v(\theta_d)\frac{|S|}{\eta}.
\end{equation}

\noindent $C_v(\theta_d)$ denotes the constant of proportionality
and displays a measure for the flow field in the vicinity of the
contact line. According to the description of Redon \etal
\cite{Red91}, the velocity in case of viscous dissipation strongly
depends on $\theta_Y$ and can also be written as

\begin{equation}
\label{Redon}
v_v\propto\frac{\gamma_{lv}}{\eta} \theta_Y^3,
\end{equation}

\noindent where the constant of proportionality accounts for the
flow singularity near the contact line. In detail, this constant
represents a logarithmic factor that includes i.a. a short-distance
cutoff and accounts for the fact that viscous dissipation diverges
near the contact line. Moreover, a small impact of the slip length
$b$ on this logarithmic factor has been experimentally found
\cite{Red91} by variation of molecular weight $M_w$ (cf.
(\ref{deGennes})) by Redon \cite{Red94}. Consequently, only in the
case of a no-slip boundary condition, the constant of
proportionality in (\ref{Redon}) and also $C_v(\theta_d)$ in
(\ref{viscousflow}) is purely independent of slip.


For growing holes of radius $R$ and dewetting velocity $V=dR/dt$,
integration of (\ref{viscousflow}) gives a linear proportionality of
the radius versus dewetting time $t$:

\begin{equation}
\label{Rpropt} R\propto t
\end{equation}

\subsubsection{Slippage - Friction at the solid/liquid interface}
\begin{figure}[t]
\begin{center}
\includegraphics[width=0.5\textwidth]{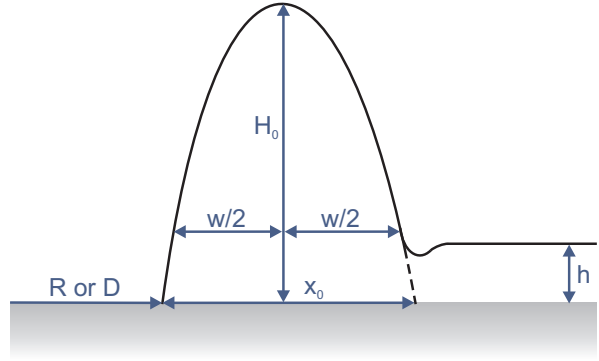}
\end{center}
\caption{Schematic representation of a rim cross-section
illustrating the rim width $w$ (usually obtained as the distance
between the three-phase contact line and the position where the rim
height is dropped to $1.1\,h$), the distance $x_0$ for a given film
thickness $h$ and hole radius $R$ (or dewetted distance $D$ of a
straight front). (Height scale in sketch is exaggerated as compared
to the lateral length scale.)} \label{graphRim2}
\end{figure}

Besides viscous dissipation within the rim, de Gennes expected a
long-chained polymer film to exhibit an exceptional amount of
slippage. For entangled polymer melts on a non-adsorbing surface,
de Gennes stated that the slip length should strongly increase
with the chain length of the polymer \cite{deG79}:

\begin{equation}
\label{deGennes} b=a\frac{N^3}{N_e^2},
\end{equation}

\noindent where $a$ is the size of the monomer, $N$ the
polymerization index and $N_e$ the entanglement length. In case of
dominating slippage, a minor contribution of viscous dissipation is
expected due to very small stresses within the rim. The
corresponding analytical model (see \cite{Bro94,Red94}) is based on
the linear friction force per unit length of the contact line given
by $F_s\propto\xi v_s$. Dissipation occurs along the distance of the
solid/liquid interface, where liquid molecules are moved over the
substrate. In general, this distance is identified as the width $w$
of the rim (cf. Fig. \ref{graphRim2}).

Consequently, $F_s\propto w$ can be assumed. If the slip length is
introduced according to Navier's model by $b=\eta/\xi$ (see
(\ref{sliplength2}) in section \ref{Navier}), the slip velocity
contribution $v_s$ is given by:

\begin{equation}
\label{slippagecontrib}
v_s=\frac{1}{3}\frac{|S|}{\eta}\frac{b}{w}.
\end{equation}

\noindent Let's consider the conservation of mass for a growing
hole of radius $R$,

\begin{equation}
\label{cm} \pi(R+x_0)^2h=2\pi Q (R+w/2),
\end{equation}

\noindent and for a straight dewetting front of dewetted distance
$D$,

\begin{equation}
\label{cm2} (D+x_0)lh=Ql,
\end{equation}

\noindent where $x_0$ stands for the distance as depicted in Fig.
\ref{graphRim2}, $Q$ denotes the cross sectional area of a rim
(approximated as a half-circle \footnote{Note that usually a dynamic
contact angle $\theta_d < 90\,^{\circ}$ at the three-phase contact
line is obtained in experiments. The difference between a segment of
a circle and the half-circle approximation is marginal and
contributes to the constant of proportionality $C_s$.}, i.e.
$Q=\pi(x_0/2)^2\approx\pi(w/2)^2$), and $l$ is the length of a
straight front, which cancels out. Based on the assumption of
self-similar growing rims and the fact that $x_0/R\ll1$ and
$x_0/D\ll1$, the width of the rim $w\propto \sqrt{h} \sqrt{R}$ for
both geometries. The constant of proportionality, however, which we
name $C_s$ in the following, depends on the shape of the rim and
furthermore on the dewetting geometry itself:

\begin{equation}
\label{w} w=C_s\sqrt{h}\sqrt{R}.
\end{equation}

\noindent Values for $C_s$ can be obtained collecting a temporal
series of AFM snapshots of rim profiles and fitting (\ref{w}) to the
measured rim width $w$ plotted versus the hole radius $R$. The
initial film thickness $h$ can also be measured by AFM. Of course,
the validity of this simplified model and especially the
approximation $w\approx x_0$ becomes questionable for asymmetric
rims and should lead to systematic differences in $C_s$. We point
out that, as described in section \ref{shapeofadewettingrim}, the
degree of asymmetry of the cross-section of the rim is strongly
correlated to the capillary number and the ratio of slip length to
film thickness. Replacing the above derived relation of rim width
$w$ and hole radius $R$ in (\ref{slippagecontrib}) gives the slip
velocity contribution in terms of the hole radius $R$ for purely
slipping liquids:

\begin{equation}
\label{slippagecontrib2}
v_s=\frac{1}{3}\frac{|S|}{\eta}\frac{b}{C_s\sqrt{h}}\frac{1}{\sqrt{R}}.
\end{equation}

\noindent Via integration, a characteristic growth law for the
radius $R$ of a dewetting hole with time $t$ is found (separation
of variables),

\begin{equation}
\label{Rpropzdt} R\propto t^{2/3}.
\end{equation}

Using the lubrication models described in section \ref{TFsection},
M\"unch compared numerical simulations of polymer melts dewetting
from hydrophobized substrates to the above described dewetting
dynamics obtained from scaling arguments (based on energy
balances) \cite{Mue051}: In the no-slip situation (mobility
$m(h)\propto h^3$), an exponent $\alpha=0.913$ is reported instead
of 1. The deviation is traced back to the fact that the
logarithmic factor in the constant of proportionality of equation
(\ref{Redon}) also depends on the width of the rim (which evolves
with time $t$). Numerical simulations of the slip-dominated case
($m(h)\propto h^2$) give $\alpha=0.661$, which captures
$\alpha=2/3$ very well.

\subsubsection{Models based on the superposition of dissipation
mechanisms} \label{section_superposition}

Considering the models derived before for pure viscous flow and for
pure slippage, a combination of these models seems to be reasonable
for situations, where viscous dissipation as well as dissipation at
the solid/liquid interface act together. Jacobs \etal proposed an
additive superposition of the corresponding velocity contributions,
i.e. $V=v_s+v_v$ \cite{Jac98}, according to the fact that both
mechanisms counteract the same driving force $|S|$. Separation of
variables leads to the implicit function:

\begin{equation}
\label{Jacobs}
t-t_0=\frac{K_v}{|S|}(R-2\frac{K_v}{K_s}\sqrt{R}+2(\frac{K_v}{K_s})^2\ln{(1+\frac{K_s\sqrt{R}}{K_v})})
\end{equation}

\noindent with

\begin{equation}
\label{kvks} K_v=\frac{\eta}{C_v(\theta_d)},
\\
K_s=\frac{3\eta}{b}\frac{w}{\sqrt{R}},
\\
\frac{K_v}{K_s}\propto b \sqrt{R}.
\end{equation}

\noindent In that notation, the velocity contributions are given by
$v_v=|S|/K_v$ and $v_s=|S|/(\sqrt{R}K_v)$. One of the preconditions
of this superposition was the successful inclusion of both border
cases: For $b\rightarrow 0$, $R\propto t$ is obtained, whereas for
$b\rightarrow \infty$, $R\propto t^{2/3}$ follows.

\begin{figure}[t]
\begin{center}
\includegraphics[width=0.8\textwidth]{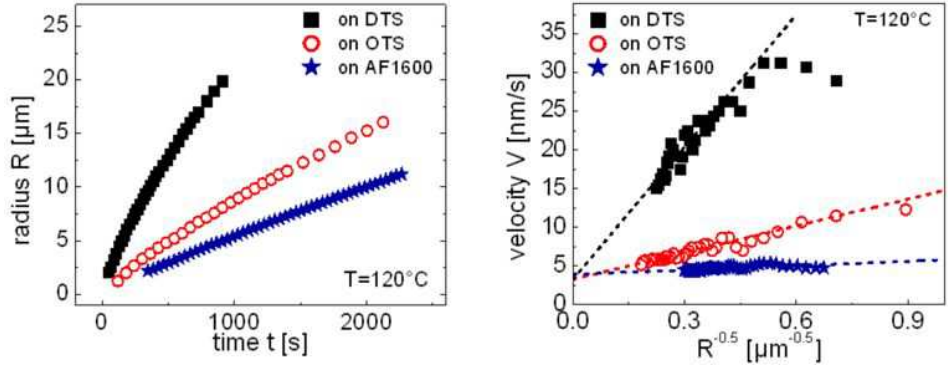}
\end{center}
\caption{Left: Optical measurement of the hole radius versus time in
130\,nm PS(13.7\,kg/mol) films on different substrates. Right: Plot
of the dewetting velocity $V$ versus $1/\sqrt{R}$. The y-axis
intercept is identical for all substrates. The difference in the
slope $K$ indicates substantial differences in slip for dewetting PS
(adapted from \cite{Fet073} and \cite{Bae08}).}
\label{graphVelocities}
\end{figure}

As a more practicable alternative for fitting this relation between
radius $R$ and time $t$ to experimental data (typical data is shown
in Fig. \ref{graphVelocities}), Fetzer and Jacobs recently proposed
a more simple way of visualizing slip effects \cite{Fet073}:
According to the additive superposition, the dewetting velocity can
be written in terms of

\begin{equation}
\label{superposition} V=v_v+\frac{K}{\sqrt{R}},
\\
K=\frac{1}{3}\frac{|S|}{\eta}\frac{b}{C_s\sqrt{h}}.
\end{equation}

\noindent which leads to a linear relationship when plotting the
dewetting velocity $v$ versus $1/\sqrt{R}$ (c.f. Fig.
\ref{graphVelocities}). Then, the y-axis intercept of the straight
line can be identified as the viscous velocity contribution $v_v$,
whereas the the slope $K$ is connected to the slip length as
illustrated in (\ref{superposition}). The validity of this model,
which assumes the linear superposition of velocity contributions,
was checked by plotting the viscous velocity contribution $v_v$
versus the reciprocal viscosity $\eta$. An excellent agreement has
been demonstrated in view of melt viscosity data obtained from
independent measurements. Moreover, the slope has been used to
calculate the slip length $b$. In the experiments, substantial
differences in slip lengths for different substrates (cf. Fig.
\ref{graphVelocities}) at identical liquid properties could be
detected \cite{Fet073}.

\subsubsection{Molecular-kinetic theory and further approaches}
Besides theoretical models based on continuum hydrodynamics (see
\ref{viscous friction}), contact line dynamics such as the spreading
of a liquid droplet on a planar surface has been analyzed according
to the so-called molecular-kinetic theory (MKT)
\cite{Bla69,deR99,Bla02}. The movement of a contact line is
described as an activated rate process, where the liquid molecules
close to the substrate jump from one potential well to another. The
resulting contact line friction coefficient is proportional i.a. to
the viscosity $\eta$ of the liquid and to the exponential of the
reversible work of adhesion $\gamma_{lv}(1+\cos{\theta_Y})/k_BT$. As
a consequence, increasing the temperature reduces the friction
coefficient and increases the velocity. In case of a squalane
droplet spreading on gold surfaces covered with different
self-assembled monolayers (SAM) of thiols, the corresponding contact
line friction coefficient turns out to increase linearly with the
chain length of the SAM \cite{Vou07}. SAM surfaces have also been
target of friction experiments using AFM tips. Barrena \etal
explained discrete changes in the frictional behavior with discrete
molecular tilts of the chains of the SAM \cite{Bar99}. Viscoelastic
deformation of the substrate at the contact line due to the normal
component of the liquid surface tension has also attracted attention
with regard to fluid dynamics on the nanoscale and energy
dissipation. The phenomenon of a reduced velocity of contact-line
displacements compared to a corresponding non-deformable substrate
(termed ''viscoelastic braking'') was extensively studied and
described by Shanahan and Carr\'{e}
\cite{Sha94,Sha95,Car95,Car96,Sha02}. In this context, Long \etal
theoretically examined the static and dynamic wetting properties of
liquids on thin rubber films \cite{Lon961} and grafted polymer
layers \cite{Lon962}. To the best of our knowledge, there is to-date
no experimental evidence of deformations and tilting on the
molecular scale in case of SAM surfaces due to retracting
contact-lines of a dewetting polymer film. However, dewetting on
soft deformable rubber surfaces has recently been shown to be
influenced by viscoelastic deformations of the substrate
\cite{Ala08}. In case of non-volatile liquids such as polymers,
evaporation into and condensation from the vapor phase can be
excluded as relevant mechanism of energy dissipation.

\subsection{Dynamics of growing holes}

Besides very early theoretical and experimental studies of thin film
rupture \cite{Vri66,Ruc74,Bro87,Sha89,Bro90,Bro921,Rei92}, one of
the first studies concerning dewetting experiments had been
published by Redon \etal \cite{Red91}. Films of alkanes and
polydimethylsiloxane (PDMS) have been prepared on different
hydrophobized silicon wafers. The authors observed a constant
dewetting velocity which was inversely proportional to the viscosity
and very sensitive to the equilibrium contact angle of the liquid on
a corresponding surface. In 1994, Redon \etal investigated the
dewetting of PDMS films of different thicknesses on silanized
hydrophobic surfaces \cite{Red94}. They found that for films larger
than 10 $\mu$m a constant dewetting velocity is found, whereas
$R(t)\propto t^{2/3}$ in films thinner than 1 $\mu$m is observed.
These results corroborate the models represented by (\ref{Rpropt})
and (\ref{Rpropzdt}) and indicate that for sufficiently thick films
($h\gg b$) viscous dissipation dominates whereas slippage becomes
relevant in case of thin films ($h<b$) \cite{Red94}.

\subsubsection{Stages of Dewetting}
While studying the growth of holes, Brochard-Wyart \etal proposed
a set of subsequent stages that can be attributed to distinct
growth laws \cite{Bro97}: Starting with the birth of the hole, the
radius of the hole is supposed to grow exponentially with time,
i.e. $R(t)\propto\exp{(t/\tau)}$, as long as the radius $R$ is
smaller than $R_c\approx \sqrt{bh}$. Experimentally, Masson \etal
\cite{Mas02} found relaxation times orders of magnitude larger
than the largest translational reptation times expected. For
$R_c<R<R_c^\prime\approx b$, the rim is formed and viscous
dissipation dominates the hole growth dynamics, i.e. $R\propto t$.
The special case of $R\gg R_c=\sqrt{bh}$ moreover gives an
analytic expression for the hole growth:

\begin{equation}
R(t)\approx \frac{|S|}{\eta}(\frac{b}{h})^{1/2}t
\end{equation}

\noindent During this linear regime, a dynamic (receding) contact
angle $\theta_d=\theta_Y/\sqrt{2}$ is formed at the three-phase
contact line. In the subsequent stage (for hole radii approximately
larger than the slip length $b$), the rim is fully established and
grows in a self-similar manner. As discussed in section \ref{viscous
friction}, this results in a characteristic $R\propto t^{2/3}$
growth law if large slip is present. That stage of self-similar
growing rims is often called ''mature'' regime. These distinct
regimes have been experimentally observed by Masson \etal
\cite{Mas02} and Damman \etal \cite{Dam03}. Moreover, a dissipation
dominated by slippage has been shown to be restricted to
sufficiently small hole radii: If the rim has accumulated a very
large amount of liquid (so that the height of the rim $H_0$ is much
larger than the slip length $b$, i.e. $H_0\gg b$), viscous
dissipation dominates again. This consequently results in a
transition to a linear growth law for the radius with respect to the
dewetting time, i.e. $R\propto t$.

In this article, we mainly focus on dewetting phenomena in terms of
growing holes or retracting straight fronts. Additionally, also the
subsequent regime influenced by the fingering instability has
attracted the interest of researchers \cite{Bro923}. Thereby, facets
such as the onset of the instability and the morphology of liquid
profiles have also shown to be sensitive to rheological properties
of thin films and the hydrodynamic boundary condition at the
solid/liquid interface \cite{Rei012,Mue052,Gab06,Kin09}.

\subsubsection{''Mature'' holes}
Recently, Fetzer \etal experimentally studied the dewetting dynamics
of ''mature'' holes in thin PS films on two different types of
hydrophobized surfaces. The aforementioned assumption of linear
superposition of viscous dissipation and slippage (see
\ref{section_superposition}) was used to gain information about the
slip contribution \cite{Fet073}. Hydrophobization was achieved by
the preparation of a dense, self-assembled monolayer of silane
molecules (silanization). In that case two different silanes have
been utilized, octadecyltrichlorosilane (OTS) and
dodecyltrichlorosilane (DTS). The slip length turns out to depend
strongly on temperature (and thereby on the melt viscosity); in
fact, the slip length decreased with increasing temperature.
Furthermore, slippage was about one order of magnitude larger on DTS
compared to OTS. The results are in good agreement with the results
for the slip length gained by the analysis of the shape of rim
profiles of the corresponding holes (see
\ref{shapeofadewettingrim}). The molecular mechanisms of slippage
are widely discussed and several models have been proposed with
regard to different experimental conditions (shear rates, polymers,
surfaces, ...). This will be discussed in the next section.

\subsection{Molecular mechanisms of slippage}
\label{shear-dependent-slippage}

For small flow velocities, slip lengths lower than expected have
been found (see e.g. \cite{Mig93}). This has been explained by the
adsorption of polymer chains at the solid surface. These adsorbed
chains are able to inhibit slippage due to entanglements with chains
in the melt. At larger shear strains, disentanglement of adsorbed
and melt chains occurs, the friction coefficient decreases rapidly
and slippage is ''switched on'' \cite{Bro92,Ajd94,Gay99,Tch05}. The
authors interpret the results as follows: Strongly adsorbed chains
stay at the solid/liquid interface followed by a coil-stretch
transition of these tethered chains. Thereby, the interaction of
anchored chains (which are stretched under high flow velocities)
with the chains in the melt are reduced and slippage is enhanced.
This has been experimentally observed by Migler \etal \cite{Mig93}
and Hervet \etal \cite{Herv03}. Thereby, the critical shear strain
depends on the molecular weight of chains attached to the surface
and on their density \cite{Leg03}. This situation is also called
''stick-slip transition'' and has moreover been reported for
pressure-controlled capillary flow of polyethylene resins
\cite{Drd95}. The authors find a characteristic molecular weight
dependence of the slip length $b\propto M_w^{3.4}$, whereas the
critical stress $\sigma_c$  for the transition described before
scales as $\sigma_c \propto M_w^{-0.5}$ \cite{Wan96}. Also
de-bonding of adsorbed chains might occur.

Molecular dynamics (MD) studies aiming to investigate the slip
length in thin, short-chained polymer films subject to planar shear
revealed a dynamic behavior of the slip length $b(\dot\gamma)$ upon
the shear rate according to
$b=b^*(1-\dot\gamma/\dot\gamma_c)^{-1/2}$, where $\gamma_c$
represents a critical shear rate \cite{Pri04}. Slippage strongly
increases as $\dot\gamma/\dot\gamma_c\rightarrow 1$. This relation
was also found for simple liquids. Additionally, the authors studied
the border case $\dot\gamma\rightarrow 0$ and especially the
molecular parameters affecting the slip length.

Moreover, a special situation has been discussed in the literature:
A thin polymer film prepared on a surface decorated with end-grafted
polymer chains of the same species. Due to entropic reasons, an
interfacial tension between identical molecules occurs and dewetting
can take place. This phenomenon is called autophobicity or
autophobic dewetting \cite{Rei00,Rei001}. Reiter and Khanna observed
that PDMS molecules slipped over grafted PDMS brushes. They
determined a slip length on the order of 10~$\mu$m for film
thicknesses between 20 and 850\,nm of the dewetting PDMS film (high
$M_w$ of 308\,kg/mol in most cases). For lower grafting densities,
slippage is reduced, indicating a deeper interpenetration of free
melt chains \cite{Rei00}.

\subsection{Impact of viscoelasticity and stress relaxation}
\label{Reiter}

As already introduced in \ref{viscoelasticity} and furthermore
theoretically described in section \ref{TFsection}, viscoelasticity
can be included in the description of thin film dynamics. In case of
thin film flow with strong slippage, we already mentioned that the
impact of viscoelasticity might be not distinguishable from slippage
effects. Numerous experimental studies have been published by
Reiter, Damman and coworkers describing the influence of
viscoelastic effects on the flow dynamics of thin polymer films.
These studies have been supported with theories by Rapha\"el and
coworkers. In the following, a snap-shot of recent results in
chronological order is presented. It reflects the progress from
simple to more and more sophisticated models and experiments. Some
of the results, however, may have tentative character, as brought up
by a very recent manuscript by Copp\'{e}e \etal (cf. section
\ref{temporalslip}).

\subsubsection{Thin film rupture}

Reiter and de Gennes \cite{Rei011} pointed out that the usual
spin-casting preparation (based on fast solvent evaporation) of
thin, long-chained polymer films may induce a cascade of distinct
states. Annealing the samples to high temperatures for times larger
than the reptation time of the polymer (depending on $M_w$ and the
temperature $T$) might induce a complete healing of the preparation
effects. Non-equilibrated conformational states and residual
stresses have shown to be capable to cause rupture of thin films
\cite{Rei05}. Thereby, the areal density of holes appearing at a
certain temperature above $T_g$ has been measured: After storage at
elevated temperatures below $T_g$, an exponential reduction of the
hole density can be observed. Vilmin and Rapha\"el were able to show
that lateral stress reduces the critical value for surface
fluctuations initiated by an anisotropic diffusion of polymer
molecules to induce the formation of holes \cite{Vil062}.

\subsubsection{Stages of distinct dewetting dynamics - Experimental results}

Reiter and co-workers furthermore studied the temporal and spatial
evolution of dewetting PS fronts on Silicon wafers covered with a
PDMS monolayer at times shorter than the relaxation (reptation) time
$\tau_{rept}$ in equilibrated bulk samples \cite{Rei05}. At the
beginning of their experiments, the dewetted distance and the rim
width $w$ increased in a logarithmic way with respect to the
dewetting time $t$, consequently $V\propto t^{-1}$ is found. They
correlated a maximum of the rim width $w$ occurring at a distinct
time $t_1$ (nearly independent on $M_w$ and significantly shorter
than $\tau_{rept}$) of the experiment to a change in dewetting
dynamics, namely $V\propto t^{-1/2}$ (i.e. $R \propto \sqrt{t}$).

Even before, Damman, Baudelet and Reiter compared the growth of a
hole and the dynamics of a dewetting straight front \cite{Dam03}.
Thereby, strong influence of the dewetting geometry on the dynamics
of early stages became obvious. Moreover, the authors correlated
their findings to the shape of the corresponding rim and clearly
identified distinct dewetting stages: At the beginning of hole
growth, capillary forces dominate the dynamics and exponential
growth is observed. In contrast to that, dewetting of a straight
front at the beginning starts almost instantaneously at a high
velocity and decreases as $V\propto t^{-1}$. As already observed by
Redon \etal \cite{Red91}, $V=$ const and $R \propto t$ due to
viscous dissipation is found for growing holes. For both dewetting
geometries, Damman \etal find a maximum of the rim width $w$ which
they interpret in terms of a transition of the shape of the rim from
very asymmetric profiles towards more symmetric rims, while the
volume still increases during dewetting. In the following
dissipation-dominated regime, $V\propto t^{-1/3}$ and $R\propto
t^{2/3}$ (slip boundary condition) or $V=$ const and $R\propto t$
(no-slip boundary condition) is found, depending on the boundary
condition for the solid/liquid interface. Afterwards, again a
constant dewetting velocity (until the ''mature rim'' regime is
reached and slippage dominates) is obtained.

Recently, Damman and Reiter determined the strain $\gamma$ from rim
shapes of dewetting fronts, which can be written according to
(\ref{Non-Newtonian}) by $\gamma=(S/h+\sigma_0)/G$, where $\sigma_0$
denotes residual stresses \cite{Dam07}. These values were one order
of magnitude larger than for equilibrated PS films
($\gamma=S/(hG_{bulk})$) and increased with increasing $M_w$.
Consequently, larger residual stresses $\sigma_0$ or smaller elastic
moduli $G$ can be responsible for this effect. Indeed, the elastic
modulus of a thin film is supposed to be smaller than its bulk value
according to the fact that larger numbers for the entanglement
length $M_e$ (for larger $M_w$) are expected (see section
\ref{confinement}). Additionally, instead of one transition time for
both, two different transitions times $t_w$ for the maximum of the
rim width and $t_v$ for the power law of the dewetting velocity (to
$V \propto t^{-1/3}$) are found for larger molecular weights. Time
$t_v$ is comparable to the reptation time $t_{rept}$ (also for
larger $M_w$). The authors interpret this as follows: After $t_v\sim
t_{rept}$, the equilibrium entanglement conformation has been
reached via interdiffusion and re-entangling of chains. Relaxation
of stress, which is correlated to the rim shape transition and
$t_w$, occurs at shorter times.

\subsubsection{Stages of distinct dewetting dynamics - Theoretical models}

These observations have motivated a theoretical model by Vilmin and
Rapha\"el \cite{Vil05}, that takes viscoelastic properties of the
liquid and slippage into account. The logarithmic increase of the
dewetted distance with time can be explained if residual stress is
regarded as an extra driving force for dewetting that has to be
added to the capillary forces. Subsequent relaxation of the stress
reduces this additional contribution. For times larger than the
reptation time $\tau_{rept}$, Vilmin and Rapha\"el predict a
constant dewetting velocity and $V \propto t^{-1/3}$ as soon as the
''mature regime'' is reached (and slippage becomes important). They
consider a simple equation for a viscoelastic film which includes
one relaxation time $\tau_1$ (and an elastic modulus $G$), but two
distinct viscosities: $\eta_0$ is a short-time viscosity which
describes the friction between monomers, and $\eta_1$ is the melt
viscosity ($\eta_1 \ll \eta_0$) governed by disentanglements of
polymer chains. Theoretically, three regimes of distinct
time-response are expected: For $t<\tau_0=\eta_0/G$, Newtonian flow
accompanied with a small viscosity $\eta_0$. Afterwards
($\tau_0<t<\tau_1=\eta_1/G$), the elastic modulus governs the
dynamics. For longer times, i.e. $t>\tau_1$, Newtonian flow is again
obtained, but with a much larger viscosity $\eta_1$. During the
first and the last regime, asymmetric rims and constant velocities
are predicted (the latter velocity of course is much smaller than
the initial velocity). The intermediate regime, governed by
viscoelastic behavior, is expected to show $V\propto t^{-1/2}$
according to the fact that the width of the rim $w$ will increase
proportional to the dewetted distance $D$, i.e. $w\propto D$ (in
contrast to the square-root dependence in case of viscous flow).
Including residual stresses into their model, Vilmin and Rapha\"el
obtained the aforementioned experimentally observed $V\propto
t^{-1}$ law instead of $V\propto t^{-1/2}$. Recently, Yang and
co-workers experimentally quantified the molecular recoiling force
stemming from non-equilibrium chain conformations. They obtained
values that are at least comparable to (or even larger than) the
dispersive driving forces \cite{Yan06}.

\subsubsection{Further remarks}

Concerning studies and models including viscoelastic effects, one
has to bear in mind that stress relaxation dynamics and relaxation
times in thin films have been shown to be different from bulk
polymer reptation values. As mentioned in section \ref{confinement}
and indicated by the characteristic relaxation times $\tau_1$ of
elastic constraints being significantly shorter than bulk values,
the entanglement length has shown to be significantly increased in
case of thin films.

Recently, Reiter and coworkers extended their studies with regard to
viscoelastic dewetting on soft, deformable substrates \cite{Ala08}.
The essential result was that transient residual stresses can cause
large elastic deformations in the substrate which almost stop
dewetting for times shorter than the relaxation time $\tau_{rept}$
of the polymer film. For times longer than $\tau_{rept}$, the
elastic behavior and the elastic trench in the deformable substrate
vanishes.

Vilmin and Rapha\"el applied their model for viscoelastic liquids
and residual stresses to the hole growth geometry \cite{Vil06}
with regard to the early stage (exponential growth for Newtonian
liquids). They discovered a very fast opening regime followed by a
slow exponential growth of the radius of the holes.

\subsection{Non-linear friction}
Up to now, concerning energy dissipation at the solid/liquid
interface, exclusively linear friction has been considered. As
described in section \ref{shear-dependent-slippage}, often not
only smooth and passive (non-adsorbing) surfaces have been
experimentally considered, but also grafted or adsorbed polymer
layers are used as a support for dewetting experiments. These
systems motivate the theoretical treatment of non-linear friction,
as described in the following subsection.

\subsubsection{Theoretical model}

To cover non-linear friction, Vilmin and Rapha\"el introduced a
friction force per unit area that is linear below a certain
transition velocity $v_{r}$ and non-linear above \cite{Vil06}:

\begin{equation}
\label{non-linear-friction} F_s=\xi v_r (v/v_r)^{1-r},\\
v>v_r,
\end{equation}

\noindent where $r$ denotes a so-called friction exponent.

\noindent Consequently, the effective (velocity-dependent) friction
coefficient $\xi_{eff}$ can be written in terms of

\begin{equation}
\label{effective-friction} \xi_{eff}(v)=\xi(v_r/v)^r
\end{equation}

\noindent and a velocity-dependent slip length $b(v)$ can be defined
as

\begin{equation}
\label{bvonv} b(v)=\eta/\xi_{eff}(v).
\end{equation}

\noindent An increasing velocity leads to a decreasing friction
coefficient and to pronounced slippage. Thereby, the friction at the
solid/liquid interface influences the power law of the velocity
decrease. Assuming non-linear friction in the intermediate regime
(which is governed by the viscoelastic behavior) gives (according to
$w \propto D$) $V \propto t^{-1/(2-r)}$. Including residual stress
$\sigma_0$ to this model leads to a formula for the maximum width of
the rim (depending on $r$ and $\sigma_0$). Experimentally, values
for $r$ between 0 (for low $M_w$) and 1 (for large $M_w$) have been
found \cite{Dam07}. The dependence of the nonlinearity of friction
upon molecular weight $M_w$ could be explained by the influence of
chain length on slippage (see (\ref{deGennes})).

\subsubsection{Variation of substrate properties}
Hamieh \etal focused on the frictional behavior of dewetting
viscoelastic PS films on PDMS-coated (irreversibly adsorbed) silicon
wafers \cite{Ham07}. Thereby, thickness (via the PDMS chain-length)
and preparation of the PDMS support (via the annealing temperature)
were varied. In summary, the observations are consistent with the
aforementioned (section \ref{Reiter}) experiments by Reiter, showing
a characteristic time $t_1$ for the change in dewetting dynamics and
rim shape (from highly asymmetric towards a more symmetric
equilibrated shape). Again, this transition is interpreted by the
Laplace pressure that overcomes elastic effects. Probing the maximum
rim width enables the authors to identify the impact of the friction
coefficient or the friction exponent $r$: If they prepare thicker
PDMS layers, the result is a larger maximum rim width. This result
can be explained in terms of a small increase of $r$. Consequently,
the velocity-dependent slip length $b(v)$ increases (see
(\ref{bvonv})): Thicker PDMS-layers lead to more slippage.
Concerning the characteristic time for stress relaxation $t_1$, no
significant influence of the preparation procedure (annealing
temperature of the PDMS coating) and thickness of the PDMS layer
have been found.

\subsection{Temporal evolution of the slip length}
\label{temporalslip} Most of the aforementioned dewetting
experiments of thin (viscoelastic) PS films are based on supports
consisting of a PDMS layer prepared onto a Si wafer. These PDMS
surfaces, which were assumed to be impenetrable by PS chains, turned
out to be less ideal. Recently, the aforementioned fast decay of the
dewetting velocity and the maximum of the rim width upon dewetting
time (ascribed to relaxation of residual stresses, see section
\ref{Reiter}), has also been observed in case of low molecular
weight PS \cite{Cop09}. Furthermore, neutron reflectometry
experiments revealed an interdiffusion of polymer chains at the
PS-PDMS interface below the PS bulk glass transition temperature.
The dewetting velocity accelerated as the brush thickness is
increased. In view of energy dissipation due to brush deformation
(pronounced by increasing thickness), this is contradictory to the
slower dewetting velocities expected in that case.

Ziebert and Rapha\"el recently investigated the temporal evolution
of the energy balance (viscous dissipation and sliding friction)
of thin film dewetting by numerical treatment, especially
concerning non-linear friction \cite{Zie092}. They point out that
both mechanisms have different time dependencies and propose that
simple scaling arguments such as the mass conservation $w \propto
D$ should be revisited. In case of non-linear friction, viscous
dissipation is even more important than sliding friction for times
larger than $\tau_1$, whereas for linear friction both mechanisms
are approx. equally important. The stage of ''mature'' rims
afterwards is again dominated by friction at the solid/liquid
interface. Using scaling arguments and numerical solutions of the
thin film model for viscoelastic liquids \cite{Vil05}, they
moreover showed that the aforementioned dewetting characteristics
(fast decay of the dewetting velocity, maximum of the rim width in
course of dewetting time) can be explained by the temporal
decrease of the slip length during the experiment instead of
stress relaxation \cite{Zie09}. Therefore, roughening of the
PS-PDMS interface (as detected by neutron reflectometry) or
potentially also an attachment of very few PS chains to the
silicon substrate might be responsible, if very low PDMS grafting
densities are prepared.

The attachment of melt chains to the substrate has been recently
considered by Reiter \etal \cite{Rei09}. They pointed out that the
driving force is reduced by a certain pull-out force for molecules
attached to the surface as they get stretched while resisting to
be pulled out. Consequently, this force (per unit length of the
contact line) can be written in terms of $F_p=\nu L f^\ast$, where
$\nu$ represents the number of surface-connected molecules, $L$
the length according to the stretching and $f^\ast$ the pull-out
force per chain.

\section{Conclusions and outlook}
To conclude, dewetting experiments can be regarded as a very
powerful tool to probe rheological thin film properties and
frictional mechanisms. The validity of the no-slip boundary
condition at the solid/liquid interface, for a long period of time
accepted as a standard approach in fluid dynamics, fails. Even for
Newtonian liquids such as polymers below their critical length for
entanglements substantial amount of slippage has been found. For
larger molecular weights, slippage can be even more pronounced if
chain entanglements come into play.

We have reviewed in this article theories and experiments
characterizing the statics and dynamics in thin liquid polymer
films, focussing on energy dissipation mechanisms occurring during
dewetting. Experiments are relatively easy to perform, since usually
no very low or very high temperatures are needed, neither are
high-speed cameras necessary. However, careful preparation,
preferably in a clean-room environment, of thin films is inevitable.
Concerning the experimental data, diligent interpretation and
consideration of all relevant parameters are of essential
importance: Molecular weight, film thickness (also with regard to
the molecular dimensions of a polymer coil in the respective melt),
dewetting temperature, melt viscosity, dewetting velocity and
capillary number, residual stresses, relaxation times and ageing
time represent parameters that are sometimes coupled and not easy to
disentangle. Their unique impact on the friction coefficient or the
slip length on a specific, ideally non-adsorbing and non-penetrable
substrate (to reduce the number of parameters of the system) is not
always easy to identify. In particular, the specific stage of
dewetting (early or mature stage) and the dewetting geometry (holes
or straight fronts) represent further facets and playgrounds for
experimentalists and theoreticians that, under careful
consideration, enable to gain insight into rheological or frictional
mechanisms. The dynamics and the morphology of the fingering
instability can provide additional access to the solid/liquid
boundary condition and the rheology.

To obtain a deeper understanding of the molecular mechanisms at the
solid/liquid interface is one of the main tasks in micro- and
nanofluidics. In this context, we like to highlight two essential
questions and possible pathways to answer them:

a) The liquid: What is the impact of the polydispersity of the
liquid on dewetting? Dewetting studies of polymer melts by adding a
second chemical component have shown to influence slippage
\cite{Bes07}. However, also the driving force is changed due to the
difference in chemical composition for different species. To
overcome this problem, the influence of chain length distribution on
dewetting can be probed by studying polymer mixtures instead of
monodisperse polymer melts \cite{Fet062}. Concerning a theoretical
approach to this question, dissipative particle dynamics (DPD)
simulations enable to locate energy loss in the rim and reveal
interesting results in case of two immiscible fluids of different
viscosity: In case of a low-viscosity layer at the solid/liquid
interface, faster dewetting dynamics is found that is attributed to
a lubrication effect, i.e. the sliding of the upper high-viscosity
layer \cite{Mer08}. Finally, these aspects lead to the fundamental
discussion, whether ''apparent'' slip, possibly induced by the
formation of a short-chained layer of low viscosity, is present.

b) The substrate: What is the impact of the molecular structure of
the substrate? As indicated before, the set of parameters concerning
the topographical and/or chemical structure of the support is large.
Besides parameters such as surface roughness and surface energy,
experiments can be performed on substrates e.g. decorated with an
amorphous coating, a self-assembled monolayer or ever grafted
polymer brushes of the same or different species. Scattering
techniques (using neutrons or X-rays) provide access to the
solid/liquid interface, complementing dewetting experiments and
confirming proposed mechanisms of slippage. Simulations based on
molecular dynamics (MD) of near-surface flows can help to compare
experimental results from dewetting studies to molecular parameters,
easily tunable in theoretical models, and structural changes
\cite{Ser08,Mue08}. Moreover, the evolution of coarse-grained
polymer brush/melt interfaces under flow has also been identified as
a potential application of MD studies \cite{Pas06,Pas09}. In the
end, all these facets help to obtain a universal picture of sliding
friction which can potentially lead to surfaces precisely tailored
for special microfluidic applications.

\section*{Acknowledgments}
The authors acknowledge financial support from the German Science
Foundation (DFG) under grant JA905/3 within the priority program
1165 \textit{''Micro- and nanofluidics''}.


\section*{References}
\begin{thebibliography}{100}

\bibitem{Squ05} Squires T M and Quake S R 2005 Microfluidics:
Fluid physics at the nanoliter scale \textit{Rev. Mod. Phys.}
\textbf{77} 977

\bibitem{Qua02} Thorsen T, Maerkl S J and Quake S R 2002 Microfluidic large-scale integration \textit{Science}
\textbf{298} 580

\bibitem{Les09} Leslie D C, Easley C J, Seker E, Karlinsey J M,
Utz M, Begley M R and Landers J P 2009 Frequency-specific flow
control in microfluidic circuits with passive elastomeric features
\textit{Nature Physics} \textbf{5} 231

\bibitem{Bea72} Bear J 1972 \textit{Dynamics of fluids in porous
media} (New York: Elsevier)

\bibitem{Vri66} Vrij A 1966 Possible mechanism for the spontaneous rupture of thin, free liquid films \textit{Discuss. Faraday
Soc.} \textbf{42} 14

\bibitem{Her98} Herminghaus S, Jacobs K, Mecke K, Bischof J, Fery A,
Ibn-Elhaj M, Schlagowski S 1998 Spinodal dewetting in liquid crystal
and liquid metal films \textit{Science} \textbf{282} 916

\bibitem{See012} Seemann R, Herminghaus S and Jacobs K 2001
Dewetting patterns and molecular forces: A reconciliation
\textit{Phys. Rev. Lett.} \textbf{86} 5534

\bibitem{Jac08} Jacobs K, Seemann R and Herminghaus S 2008
Stability and dewetting of thin liquid films \textit{Polymer Thin
Films} ed O K C Tsui and T P Russell (Singapore: World Scientific)

\bibitem{See013} Seemann R, Herminghaus S and Jacobs K 2001 Gaining
control of pattern formation of dewetting liquid films \textit{J.
Phys.: Condens. Matter} \textbf{13} 4925

\bibitem{Jac00} Jacobs K, Seemann R and Mecke K 2000 Dynamics of structure formation
in thin films: A special spatial analysis \textit{Lecture Notes in
Physics (Statistical Physics and Spatial Statistics)} ed D Stoyan
and K Mecke (Heidelberg: Springer)

\bibitem{Cra09} Craster R V and Matar O K 2009 Dynamics and
stability of thin liquid films \textit{Rev. Mod. Phys.} \textbf{81}
1131

\bibitem{Rub03} Rubinstein M and Colby R H 2003 \textit{Polymer Physics}
(New York: Oxford University Press)

\bibitem{Jon99} Jones R A L and Richards R W 1999 \textit{Polymers at surfaces and
interfaces} (Cambridge University Press)

\bibitem{deG71} De Gennes 1971 Reptation of a polymer chain in the presence of fixed
obstacles \textit{J. Chem. Phys.} \textbf{55} 572

\bibitem{Ked94} Keddie J L, Jones R A L and Cory R A 1994 Size-dependent depression of the glass transition temperature in polymer films
\textit{Europhys. Lett.} \textbf{27} 59

\bibitem{Mat00} Mattsson J, Forrest J A and B\"orgesson L 2000 Quantifying glass transition behavior in ultrathin free-standing polymer films
\textit{Phys. Rev. E} \textbf{62} 5187

\bibitem{Dal00} Dalnoki-Veress K, Forrest J A, de Gennes P-G and Dutcher J R 2000 Glass transition reductions in thin freely-standing polymer films :
A scaling analysis of chain confinement effects \textit{J. Phys. IV
France} \textbf{10} 221

\bibitem{Her01} Herminghaus S, Jacobs K and Seemann R 2001 The glass transition of thin polymer films: Some questions, and a possible answer
\textit{Europ. Phys. J. E} \textbf{5} 531

\bibitem{Ked95} Keddie J L and Jones R A L 1995 Glass transition behavior in ultrathin polystyrene films \textit{J. Isr. Chem. Soc.} \textbf{35} 21

\bibitem{Fry01} Fryer D S, Peters R D, Kim E J, Tomaszewski J E,
De Pablo J J and Nealey P F 2001 Dependence of the glass transition
temperature of polymer films on interfacial energy and thickness
\textit{Macromolecules} \textbf{34} 5627

\bibitem{Her03} Herminghaus S, Jacobs K and Seemann R 2003
Viscoelastic dynamics of polymer thin films and surfaces
\textit{Eur. Phys. J. E} \textbf{12} 101

\bibitem{Kaw01} Kawana S and Jones R A L 2001 Character of the
glass transition in thin supported polymer films \textit{Phys. Rev.
E} \textbf{63} 021501

\bibitem{Her04} Herminghaus S, Seemann R, and Landfester K 2004 Polymer surface melting mediated by capillary
waves \textit{Phys. Rev. Lett.} \textbf{93} 017801

\bibitem{See06} Seemann R, Jacobs K, Landfester K, and Herminghaus S
2006 Freezing of polymer thin films and surfaces: The small
molecular weight puzzle \textit{J. polym. Sci. B} \textbf{44} 2968

\bibitem{Si05} Si L, Massa M V, Dalnoki-Veress K, Brown H R and
Jones R A L 2005 Chain entanglement in thin freestanding polymer
films \textit{Phys. Rev. Lett.} \textbf{94} 127801

\bibitem{Rau05} Rauscher M, M\"unch A, Wagner B and Blossey R 2005
A thin-film equation for viscoelastic liquids of Jeffreys type
\textit{Eur. Phys. J. E} \textbf{17} 373

\bibitem{Blo06} Blossey R, M\"unch A, Rauscher M and Wagner B 2006
Slip vs. viscoelasticity in dewetting thin films \textit{Eur. Phys.
J. E} \textbf{20} 267

\bibitem{Mue06} M\"unch A, Wagner B, Rauscher M and Blossey R 2006
A thin-film model for corotational Jeffreys fluids under strong slip
\textit{Eur. Phys. J. E} \textit{20} 365

\bibitem{Nij07} Te Nijenhuis K, McKinley G H, Spiegelberg S, Barnes H A, Aksel N, Heymann L, Odell J A 2007
Non-Newtonian flows \textit{Springer Handbook of experimental fluid
mechanics} ed C Tropea, A L Yarin and J F Foss (Berlin, Heidelberg,
New York: Springer)

\bibitem{Mue05} M\"unch A, Wagner B A and Witelski T P 2005
Lubrication models with small to large slip lengths \textit{J. Eng.
Math.} \textbf{53} 359

\bibitem{Nav23} Navier C L M H 1823 M\'{e}moire sur les lois du mouvement des fluids \textit{Mem. Acad. Sci. Inst. Fr.} \textbf{6}, 389 and 432

\bibitem{Lau07} Lauga E, Brenner M P and Stone H A 2007
Microfluidics: The no-slip boundary condition \textit{Springer
Handbook of experimental fluid mechanics} ed C Tropea, A L Yarin and
J F Foss (Berlin, Heidelberg, New York: Springer)

\bibitem{Net05} Neto C, Evans D R, Bonaccurso E, Butt H-J and Craig
V S J 2005 Boundary slip in Newtonian liquids: A review of
experimental studies \textit{Rep. Prog. Phys.} \textbf{68} 2859

\bibitem{Boc07} Bocquet L and Barrat J-L 2007 Flow boundary
conditions from nano- to micro-scales \textit{Soft Matter}
\textbf{3} 685

\bibitem{Jol06} Joly L, Ybert C and Bocquet L 2006 Probing the
nanohydrodynamics at liquid/solid interfaces using thermal motion
\textit{Phys. Rev. Lett.} \textbf{96} 046101

\bibitem{Bar991} Barrat J-L and Bocquet L 1999 Large slip effect at a
nonwetting fluid-solid interface \textit{Phys. Rev. Lett.}
\textbf{82} 4671

\bibitem{Pit00} Pit R, Hervet H and L\'{e}ger L 2000 Direct
experimental evidence of slip in hexadecane: Solid interfaces
\textit{Phys. Rev. Lett.} \textbf{85} 980

\bibitem{Cot02} Cottin-Bizonne C, Jurine S, Baudry J, Crassous J,
Restagno F and Charlaix E 2002 Nanorheology: An investigation of the
boundary condition at hydrophobic and hydrophilic interfaces
\textit{Eur. Phys. J. E} \textbf{9} 47

\bibitem{Leg03} Leger L 2003 Friction mechanisms and interfacial
slip at fluid-solid interfaces \textit{J.Phys.: Condens. Matter}
\textbf{15} S19

\bibitem{Zhu02} Zhu Y and Granick S 2002 Limits of the hydrodynamic
no-slip boundary condition \textit{Phys. Rev. Lett.} \textbf{88}
106102

\bibitem{Sch06} Schmatko T, Hervet H and L\'{e}ger L 2006 Effect
of nano-scale roughness on slip at the wall of simple fluids
\textit{Langmuir} \textbf{22} 6843

\bibitem{Kun07} Kunert C and Harting J 2007 Roughness induced
boundary slip in microchannel flows \textit{Phys. Rev. Lett.}
\textbf{99} 176001

\bibitem{Cot03} Cottin-Bizonne C, Barrat J-L, Bocquet L and
Charlaix E 2003 Low-friction flows of liquid at nanopatterned
interfaces \textit{Nat. Mater.} \textbf{2} 237

\bibitem{Pri06} Priezjev N V and Troian S M 2006 Influence of
periodic wall roughness on the slip behaviour at liquid/solid
interfaces: Molecular-scale simulations versus continuum predictions
\textit{J. Fluid Mech.} \textbf{554} 25

\bibitem{Jos06} Joseph P, Cottin-Bizonne C, Beno\^{\i}t J-M, Ybert
C, Journet C, Tabeling P and Bocquet L 2006 Slippage of water past
superhydrophobic carbon nanotube forests in microchannels
\textit{Phys. Rev. lett.} \textbf{97} 156104

\bibitem{Ybe07} Ybert C, Barentin C, Cottin-Bizonne C, Joseph P and
Bocquet L 2007 Achieving large slip with superhydrophobic surfaces:
Scaling laws for generic geometries \textit{Phys. Fluids}
\textbf{19} 123601

\bibitem{Ste07} Steinberger A, Cottin-Bizonne C, Kleimann P and
Charlaix E 2007 High friction on a bubble mattress \textit{Nature
Materials} \textbf{6} 665

\bibitem{Sch05} Schmatko T, Hervet H and L\'{e}ger L 2005 Friction and
slip at simple fluid/solid interfaces: The roles of the molecular
shape and the solid/liquid interaction \textit{Phys. Rev. Lett.}
\textbf{94} 244501

\bibitem{Pri05} Priezjev N V, Darhuber A A and Troian S M 2005
Slip behavior in liquid films on surfaces of patterned wettability:
Comparison between continuum and molecular dynamics simulations
\textit{Phys. Rev. E} \textbf{71} 041608

\bibitem{Hei07} Heidenreich S, Ilg P and Hess S 2007 Boundary
conditions for fluids with internal orientational degrees of
freedom: Apparent velocity slip associated with the molecular
alignment \textit{Phys. Rev. E} \textbf{75} 066302

\bibitem{Cho04} Cho J-H J, Law B M and Rieutord F 2004
Dipole-dependent slip of Newtonian liquids at smooth solid
hydrophobic surfaces \textit{Phys. Rev. Lett.} \textbf{92} 166102

\bibitem{deG02} De Gennes P G 2002 On fluid/wall slippage
\textit{Langmuir} \textbf{18} 3413

\bibitem{Hua08} Huang D M, Sendner C, Horinek D, Netz R R and
Bocquet L 2008 Water slippage versus contact angle: A
quasiuniversial relationship \textit{Phys. Rev. Lett.} \textbf{101}
226101

\bibitem{Ste03} Steitz R, Gutberlet T, Hauss T, Kl\"osgen B, Krastev
R, Schemmel S, Simonsen A C, Findenegg G H 2003 Nanobubbles and
their precursor layer at the interface of water against a
hydrophobic substrate \textit{Langmuir} \textbf{19} 2409

\bibitem{Dos05} Doshi D A, Watkins E B, Israelachvili J N and
Majewski J 2005 Reduced water density at hydrophobic surfaces:
Effect of dissolved gases \textit{PNAS} \textbf{102} 9458

\bibitem{Mez06} Mezger M, Reichert H, Schr\"oder S, Okasinski J,
Schr\"oder H, Dosch H, Palms D, Ralston J and Honkim\"aki V 2006
High-resolution in situ x-ray study of the hydrophobic gap at the
water-octadecyl-trichlorosilane interface \textit{PNAS} \textbf{103}
18401

\bibitem{Mac07} Maccarini M, Steitz R, Himmelhaus M, Fick J, Tatur
S, Wolff M, Grunze M, Jane$\check{c}$ek, Netz R R 2007 Density
depletion at solid-liquid interfaces: A neutron reflectivity study
\textit{Langmuir} \textbf{23} 598

\bibitem{Tyr01} Tyrrell J W G and Attard P 2001 Images of
nanobubbles on hydrophobic surfaces and their interactions
\textit{Phys. Rev. Lett.} \textbf{87} 176104

\bibitem{Tre04} Tretheway D C and Meinhart C D 2004 A generating
mechanism for apparent fluid slip in hydrophobic microchannels
\textit{Phys. Fluids} \textbf{16} 1509

\bibitem{Poy06} Poynor A, Hong L, Robinson I K, Granick S, Zhang Z
and Fenter P A 2006 How water meets a hydrophobic surface
\textit{Phys. Rev. Lett.} \textbf{97} 266101

\bibitem{Hen09} Hendy S C and Lund N J 2009 Effective slip length
for flows over surfaces with nanobubbles: The effect of finite slip
\textit{J. Phys.: Condens. Matter} \textbf{21} 144202

\bibitem{Oro97} Oron A, Davis S H and Bankoff S G 1997 Long-scale evolution of thin liquid films \textit{Rev.
Mod. Phys.} \textbf{69} 931

\bibitem{Kar04} Kargupta K, Sharma A and Khanna R 2004
Instability, dynamics, and morphology of thin slipping films
\textit{Langmuir} \textbf{20} 244

\bibitem{Fet07} Fetzer R, M\"unch A, Wagner B, Rauscher M and Jacobs
K 2007 Quantifying hydrodynamic slip: A comprehensive analysis of
dewetting profiles \textit{Langmuir} \textbf{23} 10559

\bibitem{Blo08} Blossey R 2008 Thin film rupture and polymer flow
\textit{Phys. Chem. Chem. Phys.} \textbf{10} 5177

\bibitem{Isr92} Israelachvili J 1992 \textit{Intermolecular and surface
forces} 2nd edition (New York: Academic Press)

\bibitem{Bec03} Becker J, Gr\"un G, Seemann R, Mantz H, Jacobs K,
Mecke K R and Blossey R 2003 Complex dewetting scenarios captured by
thin-film models \textit{Nature Materials} \textbf{2} 59

\bibitem{Rau08} Rauscher M, Blossey R, M\"unch A and Wagner B 2008
Spinodal dewetting of thin films with large interfacial slip:
Implications from the dispersion relation \textit{Langmuir}
\textbf{24} 12290

\bibitem{Fet072} Fetzer R, Rauscher M, Seemann R, Jacobs K and Mecke
K 2007 Thermal noise influences fluid flow in thin films during
spinodal dewetting \textit{Phys. Rev. Lett.} \textbf{99} 114503

\bibitem{See01} Seemann R, Herminghaus S and Jacobs K 2001
Shape of a liquid front upon dewetting \textit{Phys. Rev. Lett.}
\textbf{87} 196101

\bibitem{Rei01} Reiter G 2001 Dewetting of highly elastic thin
polymer films \textit{Phys. Rev. Lett.} \textbf{87} 186101

\bibitem{Dam03} Damman P, Baudelet N and Reiter G 2003 Dewetting near the glass transition:
Transition from a capillary force dominated to a dissipation
dominated regime \textit{Phys. Rev. Lett.} \textbf{91} 216101

\bibitem{Net03-1} Neto C, Jacobs K, Seemann R, Blossey R, Becker J and Gr\"un G 2003 Correlated dewetting patterns in thin polystyrene
films \textit{J. Phys.: Condens. Matter} \textbf{15} 421

\bibitem{Net03-2} Neto C, Jacobs K, Seemann R, Blossey R, Becker J and Gr\"un G 2003 Satellite hole formation during dewetting: Experiment and simulation
\textit{J. Phys.: Condens. Matter} \textbf{15} 3355

\bibitem{Fet05} Fetzer R, Jacobs K, M\"unch A, Wagner B and
Witelski T P 2005 New slip regimes and the shape of dewetting thin
liquid films \textit{Phys. Rev. Lett.} \textbf{95} 127801

\bibitem{Fet06} Fetzer R, Rauscher M, M\"unch A, Wagner B A and
Jacobs K 2006 Slip-controlled thin film dynamics \textit{Europhys.
Lett.} \textbf{75} 638

\bibitem{Bae09} B\"aumchen O, Fetzer R, M\"unch A, Wagner B and Jacobs K 2009 Comprehensive analysis of dewetting profiles to quantify hydrodynamic slip
\textit{IUTAM Symp. on Adv. in Micro- and Nanofluidics} ed M Ellero,
X Hu, J Fr\"ohlich and N Adams (Springer)

\bibitem{deG85} De Gennes P G 1985 Wetting: Statics and dynamics
\textit{Rev. Mod. Phys.} \textbf{57} 827

\bibitem{Fru38} Frumkin A N 1938 On the wetting phenomena and attachment of bubbles I \textit{J. Phys. Chem. USSR}
\textbf{12} 337

\bibitem{Bro94} Brochard-Wyart F, De Gennes P-G, Hervet H and Redon C 1994 Wetting and slippage of polymer melts on semi-ideal surfaces
\textit{Langmuir} \textbf{10} 1566

\bibitem{Red91} Redon C, Brochard-Wyart F and Rondelez F 1991
Dynamics of dewetting \textit{Phys. Rev. Lett.} \textbf{66} 715

\bibitem{Red94} Redon C, Brzoska J B and Brochard-Wyart F Dewetting and slippage of microscopic polymer films 1994 \textit{Macromolecules} \textbf{27} 468

\bibitem{deG79} De Gennes P G 1979 Ecoulements viscom\'{e}triques de polym\`{e}res enchev\^{e}tr\'{e}s
\textit{C. R. Acad. Sci. B} \textbf{288} 219

\bibitem{Mue051} M\"unch A 2005 Dewetting rates of thin liquid films
\textit{J. Phys.: Condens. Matter} \textbf{17} S309

\bibitem{Jac98} Jacobs K, Seemann R, Schatz G and Herminghaus S 1998 Growth
of holes in liquid films with partial slippage \textit{Langmuir}
\textbf{14} 4961

\bibitem{Fet073} Fetzer R and Jacobs K 2007 Slippage of newtonian
liquids: Influence on the dynamics of dewetting thin films
\textit{Langmuir} \textbf{23} 11617

\bibitem{Bae08} B\"aumchen O, Jacobs K and Fetzer R 2008 Probing
slippage and flow dynamics of thin dewetting polymer films
\textit{Proc. Eur. Conf. on Microfluidics (Bologna) Dec 10-12, 2008}

\bibitem{Bla69} Blake T D and Haynes J M 1969 Kinetics of liquid/liquid displacement
\textit{J. Colloid Interface Sci.} \textbf{30} 421

\bibitem{deR99} De Ruijter M J, Blake T D and De Coninck J 1999
Dynamic wetting studied by molecular modeling simulations of droplet
spreading \textit{Langmuir} \textbf{15} 7836

\bibitem{Bla02} Blake T D and De Coninck J 2002 The influence of
solid/liquid interactions on dynamic wetting \textit{J. Adv. Colloid
Interface Sci.} \textbf{96} 21

\bibitem{Vou07} Vou\'{e} M, Rioboo R, Adao M H, Conti J, Bondar A
I, Ivanov D A, Blake T D and De Coninck J 2007 Contact-line friction
of liquid drops on self-assembled monolayers: Chain-length effects
\textit{Langmuir} \textbf{23} 4695

\bibitem{Bar99} Barrena E, Kopta S, Ogletree D F, Charych D H and
Salmeron M 1999 Relationship between friction and molecular
structure: Alkylsilane lubricant films under pressure \textit{Phys.
Rev. Lett.} \textbf{82} 2880

\bibitem{Sha94} Shanahan M E R and Carr\'{e} A 1994 Anomalous
spreading of liquid drops on an elastomeric surface
\textit{Langmuir} \textbf{10} 1647

\bibitem{Sha95} Shanahan M E R and Carr\'{e} A 1995 Viscoelastic
dissipation in wetting and adhesion phenomena \textit{Langmuir}
\textbf{11} 1396

\bibitem{Car95} Carr\'{e} A and Shanahan M E R 1995 Influence of the
''wetting ridge'' in dry patch formation \textit{Langmuir}
\textbf{11} 3572

\bibitem{Car96} Carr\'{e} A, Gastel J-C and Shanahan M E R 1996
viscoelastic effects in the spreading of liquids \textit{Nature}
\textbf{379} 432

\bibitem{Sha02} Shanahan M E R and Carr\'{e} A 2002 Spreading and dynamics
of liquid drops involving nanometric deformations on soft substrates
\textit{Colloids Surf. A} \textbf{206} 115

\bibitem{Lon961} Long D, Ajdari A and Leibler L 1996 Static and
dynamic wetting properties of thin rubber films \textit{Langmuir}
\textbf{12} 5221

\bibitem{Lon962} Long D, Ajdari A and Leibler L 1996 How do grafted
polymer layers alter the dynamics of wetting \textit{Langmuir}
\textbf{12} 1675

\bibitem{Ala08} Al Akhrass S, Reiter G, Hou S Y, Yang M H, Chang Y
L, Chang F C, Wang C F and Yang A C-M 2008 Viscoelastic thin polymer
films under transient residual stresses: Two-stage dewetting on soft
substrates \textit{Phys. Rev. Lett.} \textbf{100} 178301

\bibitem{Ruc74} Ruckenstein E and Jain R K 1974 Spontaneous rupture of thin liquid
films \textit{J. Chem. Soc. Faraday Trans. II} 132

\bibitem{Bro87} Brochard-Wyart F, Di Meglio J-M and Qu\'{e}r\'{e} D
1987 Dewetting \textit{C. R. Acad. Sc. Paris S\'{e}rie II}
\textbf{304} 553

\bibitem{Sha89} Sharma A and Ruckenstein E 1989 Dewetting of solids by the formation of holes in macroscopic liquid films
\textit{J. Coll. Int. Sci.} \textbf{133} 358

\bibitem{Bro90} Brochard-Wyart F and Daillant J 1990 Drying of
solids wetted by thin liquid films \textit{Can. J. Phys.}
\textbf{68} 1084

\bibitem{Bro921} Brochard-Wyart F, Redon C and Skykes C 1992
Dewetting of ultrathin liquid films \textit{C.R. Acad. Sci. Paris
S\'{e}rie II} \textbf{314} 19

\bibitem{Rei92} Reiter G 1992 Dewetting of thin polymer films
\textit{Phys. Rev. Lett.} \textbf{68} 75

\bibitem{Bro97} Brochard-Wyart F, Debregeas G, Fondecave R and Martin P 1997 Dewetting of supported viscoelastic polymer films:  Birth of rims
\textit{Macromolecules} \textbf{30} 1211

\bibitem{Mas02} Masson J-L and Green P F 2002 Hole formation in
thin polymer films: A two-stage process \textit{Phys. Rev. Lett.}
\textbf{88} 205504

\bibitem{Bro923} Brochard-Wyart F and Redon C 1992 Dynamics of rim
instabilities \textit{Langmuir} \textbf{8} 2324

\bibitem{Rei012} Reiter G and Sharma A 2001 Auto-optimization of dewetting
rates by rim instabilities in slipping polymer films \textit{Phys.
Rev. Lett.} \textbf{87} 166103

\bibitem{Mue052} M\"unch A and Wagner B 2005 Contact-line
instability of dewetting thin films \textit{Physica D} \textbf{209}
178

\bibitem{Gab06} Gabriele S, Sclavons S, Reiter G and Damman P 2006
Disentanglement time of polymers determines the onset of rim
instabilities in dewetting \textit{Phys. Rev. Lett.} \textbf{96}
156105

\bibitem{Kin09} King J R, M\"unch A and Wagner B 2009 Linear stability analysis
of a sharp-interface model for dewetting thin films \textit{J. Eng.
Math.} \textbf{63} 177

\bibitem{Mig93} Migler K B, Hervet H and Leger L 1993 Slip
transition of a polymer melt under shear stress \textit{Phys. Rev.
Lett.} \textbf{70} 287

\bibitem{Bro92} Brochard F and de Gennes P G 1992 Shear-dependent
slippage at a polymer/solid interface \textit{Langmuir} \textbf{8}
3033

\bibitem{Ajd94} Ajdari A, Brochard-Wyart F, De Gennes P-G, Leibler
L, Viovy J-L and Rubinstein M 1994 Slippage of an entangled polymer
melt on a grafted surface \textit{Physica A} \textbf{204} 17

\bibitem{Gay99} Gay C 1999 New concepts for the slippage of an
entangled polymer melt at a grafted solid interface \textit{Eur.
Phys. J. B} \textbf{7} 251

\bibitem{Tch05} Tchesnokov M A, Molenaar J, Slot J J M and
Stepanyan R 2005 A molecular model for cohesive slip at polymer
melt/solid interfaces \textit{J. Chem. Phys.} \textbf{122} 214711

\bibitem{Herv03} Hervet H and Leger L 2003 Flow with slip at the
wall: From simple to complex fluids \textit{C. R. Physique}
\textbf{4} 241

\bibitem{Drd95} Drda PP and Wang S-Q 1995 Stick-slip transition at
polymer melt/solid interfaces \textit{Phys. Rev. Lett.} \textbf{75}
2698

\bibitem{Wan96} Wang S-Q and Drda P A 1996 Stick-slip transition
in capillary flow of polyethylene. 2. Molecular-weight dependence
and low-temperature anomaly \textit{Macromolecules} \textbf{29} 4115

\bibitem{Pri04} Priezjev N V and Troian S M 2004 Molecular origin
and dynamic behavior of slip in sheared polymer films \textit{Phys.
Rev. Lett.} \textbf{92} 018302

\bibitem{Rei00} Reiter G and Khanna R 2000 Kinetics of autophobic
dewetting of polymer films \textit{Langmuir} \textbf{16} 6351

\bibitem{Rei001} Reiter G and Khanna R 2000 Negative excess
interfacial entropy between free and end-grafted chemically
identical polymers \textit{Phys. Rev. Lett.} \textbf{85} 5599

\bibitem{Rei011} Reiter G and De Gennes P-G 2001 Spin-cast, thin,
glassy polymer films: Highly metastable forms of matter \textit{Eur.
Phys. J. E} \textbf{6} 25

\bibitem{Rei05} Reiter G, Hamieh M, Damman P, Sclavons S, Gabriele
S, Vilmin T and Rapha\"el E 2005 Residual stresses in thin polymer
films cause rupture and dominate early stages of dewetting
\textit{Nature Materials} \textbf{4} 754

\bibitem{Vil062} Vilmin T and Rapha\"el E 2006 Dynamic instability
of thin viscoelstic films under lateral stress \textit{Phys. Rev.
Lett.} \textbf{97} 036105

\bibitem{Dam07} Damman P, Gabriele S, Copp\'{e}e S, Desprez S,
Villers D, Vilmin T, Rapha\"{e}l E, Hamieh M, Al Akhrass S and
Reiter G 2007 Relaxation of residual stress and reentanglement of
polymers in spin-coated films \textit{Phys. Rev. Lett.} \textbf{99}
036101

\bibitem{Vil05} Vilmin T and Rapha\"el E 2005 Dewetting of thin
viscoelastic polymer films on slippery substrates \textit{Europhys.
Lett.} \textbf{72} 781

\bibitem{Yan06} Yang M H, Hou S Y, Chang Y L and Yang A C-M 2006
Molecular recoiling in polymer thin film dewetting \textit{Phys.
Rev. Lett.} \textbf{96} 066105

\bibitem{Vil06} Vilmin T and Rapha\"{e}l 2006 Dewetting of thin
polymer films \textit{Eur. Phys. J. E} \textbf{21} 161

\bibitem{Ham07} Hamieh M, Al Akhrass S, Hamieh T, Damman P, Gabriele S,
Vilmin T, Rapha\"el E and Reiter G 2007 Influence of substrate
properties on the dewetting dynamics of viscoelastic polymer films
\textit{J. of Adhes.} \textbf{83} 367

\bibitem{Cop09} Copp\'{e}e S, Gabriele S, Jonas A M, Jestin J,
Damman P 2009 Influence of chain interdiffusion between immiscible
polymers on dewetting dynamics \textit{arXiv:0904.1675v1}

\bibitem{Zie092} Ziebert F and Rapha\"el E 2009 Dewetting dynamics of
stressed viscoelastic thin polymer films \textit{Phys. Rev. E}
\textbf{79} 031605

\bibitem{Zie09} Ziebert F and Rapha\"{e}l E 2009 Dewetting of thin
polymer films: Influence of interface evolution \textit{Europhys.
Lett.} \textbf{86} 46001

\bibitem{Rei09} Reiter G, Al Akhrass S, Hamieh M, Damman P,
Gabriele S, Vilmin T and Rapha\"el E 2009 Dewetting as an
investigative tool for studying properties of thin polymer films
\textit{Eur. Phys. J. Special Topics} \textbf{166} 165

\bibitem{Bes07} Besancon B M and Green P F 2007 Dewetting dynamics
in miscible polymer-polymer thin films mixtures \textit{J. Chem.
Phys.} \textbf{126} 224903

\bibitem{Fet062} Fetzer R 2006 \textit{Einfluss von Grenzfl\"achen auf die
Fluidik d\"unner Polymerfilme} (Berlin: Logos)

\bibitem{Mer08} Merabia S and Avalos J B 2008 Dewetting of a
stratified two-component liquid film on a solid substrate
\textit{Phys. Rev. Lett.} \textbf{101} 208304

\bibitem{Ser08} Servantie J and M\"uller M 2008 Temperature
dependence of the slip length in polymer melts at attractive
surfaces \textit{Phys. Rev. Lett.} \textbf{101} 026101

\bibitem{Mue08} M\"uller M, Pastorino C and Servantie J 2008 Flow, slippage
and a hydrodynamic boundary condition of polymers at surfaces
\textit{J. Phys.: Condens. Matter} \textbf{20} 494225

\bibitem{Pas06} Pastorino C, Binder K, Kreer T and M\"uller M 2006
Static and dynamic properties of the interface between a polymer
brush and a melt of identical chains \textit{J. Chem. Phys.}
\textbf{124} 064902

\bibitem{Pas09} Pastorino C, Binder K and M\"uller M 2009
Coarse-grained description of a brush/melt interface in equilibrium
and under flow \textit{Macromolecules} \textbf{42} 401

\end{thebibliography}
\end{document}